\documentclass[12pt,english,longbibliography,aps,nofootinbib,preprint]{revtex4-1}
\usepackage{hyperref}
\usepackage{setspace}
\usepackage{mathptmx}
\usepackage{graphicx}

\usepackage[T1]{fontenc}
\usepackage[latin9]{inputenc}
\usepackage[letterpaper]{geometry}
\usepackage{tikz}
\usepackage{parskip}
\usepackage{booktabs}
\usepackage{amsmath}
\usepackage{amssymb}
\usepackage{amsthm}
\usepackage{cancel}
\usepackage{babel}
\usepackage{xcolor}
\usepackage{mathtools}
\newtheorem{theorem}{Theorem}[section]
\newtheorem{corollary}[theorem]{Corollary}
\newtheorem{lemma}[theorem]{Lemma}

\theoremstyle{definition}

\begin{document}

\title{Ground State Excitations and Energy Fluctuations in Short-Range Spin Glasses}
\begin{abstract}

We study the stability of ground states in the Edwards-Anderson Ising spin glass in dimensions two and higher against perturbations of a single coupling. After reviewing the concepts of critical droplets, flexibilities and metastates, we show that, in any dimension, a certain kind of critical droplet with space-filling (i.e., positive spatial density) boundary does not exist in ground states generated by coupling-independent boundary conditions.  Using this we show that if incongruent ground states exist in any dimension, the variance of their energy difference restricted to finite volumes scales proportionally to the volume. This in turn is used to prove that a metastate generated by (e.g.) periodic boundary conditions is unique and supported on a single pair of spin-reversed ground states in two dimensions. We further show that any excitation above an infinite-volume ground state with a positive-density interface must have the property that its energy difference with the ground state in a restricted volume diverges as the square root of the restricted volume.

\end{abstract}
\author{C.M.~Newman}
\affiliation{Courant Institute of Mathematical Sciences, New York University, New York, NY 10012 USA}
\affiliation{NYU-ECNU Institute of Mathematical Sciences at NYU Shanghai, 3663 Zhongshan Road North, Shanghai, 200062, China}
\author{D.L.~Stein}
\affiliation{Department of Physics and Courant Institute of Mathematical Sciences, New York University, New York, NY 10012 USA}
\affiliation{NYU-ECNU Institutes of Physics and Mathematical Sciences at NYU Shanghai, 3663 Zhongshan Road North, Shanghai, 200062, China}
\affiliation{Santa Fe Institute, 1399 Hyde Park Rd., Santa Fe, NM USA 87501}

\maketitle

\section{Introduction}

Although the thermodynamic behavior of mean-field spin glasses is now well-understood~\cite{CMMPRSZ23}, that of finite-dimensional spin glasses with short-range
interactions remains controversial.  Two of the most important open questions concern the zero-temperature properties of the spin glass:  how many distinct (i.e., not 
related via a global spin flip) ground states are present in the thermodynamic limit, and what is the nature of their lowest-energy large-lengthscale excitations?
 The answers to these questions are important not only in determining the thermal properties of the spin glass phase at low but nonzero temperatures, but are also
 relevant to certain dynamical questions such as the nonequilibrium evolution of a spin glass following a deep quench~\cite{NS99a,NS99c,NS08,JRY21}.
 
 In previous work~\cite{NS22,NS24c} the authors studied the stability of spin glass ground states with respect to perturbations of a single coupling, and identified
 a particular type of instability, called a space-filling critical droplet (to be described below), which played a central role in determining which of several proposed
 scenarios for the spin glass ground state~\cite{NS03a,NS22,NRS23b} describes its actual behavior.  In this paper we show that such instabilities do not exist in any dimension,
 with the consequence that fluctuations in the energy difference, restricted to a subvolume, between two infinite-volume ground states diverge proportionally to the volume. 
 (We consider only continuous coupling distributions and ground states that are limits of an infinite sequence of finite-volume ground states generated with coupling-independent boundary
 conditions, such as free, periodic, or fixed.) This leads to several results, including a proof that in two dimensions there is only a single pair of spin-reversed infinite-volume
 ground states, and that in any dimension excitations above infinite-volume ground states which are space-filling (i.e., differ from the ground state on a positive density of edges) must have an energy difference with the ground state which diverges with the restricted volume considered.

The paper is organized as follows.   Sections~\ref{subsec:gs}~--~\ref{subsec:meta} provide a review of basic definitions and relevant features of spin glass ground states, critical droplets and their flexibilities, and metastates. Sections~\ref{subsec:properties} and~\ref{subsec:types} review previously obtained results on the properties and types of critical droplets. Readers who are already familiar with these concepts can skip to Section~\ref{sec:continuity} and refer to Section~\ref{sec:review} as needed. 

Section~\ref{sec:continuity} examines the behavior of the flexibility when a coupling passes through its critical value (defined in Sect.~\ref{subsec:cd}) and proves a result regarding its continuity needed in later sections. Section~\ref{sec:sfcd} focuses on space-filling critical droplets, the main object of interest in this paper.  A study of possible scenarios that can give rise to such droplets culminates in Theorem~\ref{thm:noexist}, which asserts that space-filling critical droplets do not occur in ground states in the support of a translation-covariant metastate in any dimension.  This is one of the central results of this paper. 

The remaining sections examine the consequences of Theorem~\ref{thm:noexist}. Section~\ref{sec:energyflucs} shows how the absence of space-filling critical droplets allows for the extension to zero temperature of previously obtained results for spin glasses at positive temperature~\cite{NS24a,NS24b,NS24c}. This extension leads to Theorem~\ref{thm:cd}, the paper's second main result, which asserts  that in any dimension the scale of energy fluctuations between two incongruent spin glass ground states diverges with the volume in which the fluctuations are measured. Finally, Section~\ref{sec:discussion} derives two consequences of the results obtained in earlier sections. In Section~\ref{subsec:2D} we present a proof that a translation-covariant metastate of the Edwards-Anderson~\cite{EA75} Ising spin glass is supported on a single spin-reversed ground state pair in two dimensions (see Theorems~\ref{thm:single} and~{\ref{thm:unique}).  In Section~\ref{subsec:RSB} we consider the possibility of large-lengthscale (and therefore thermodynamically relevant) excitations above a spin glass ground state such that in infinite volume the excitation/ground state interface is space-filling and has an energy scale diverging more slowly than the square root of the volume.  As we will see in Section~\ref{subsec:RSB}, such interfaces cannot exist in any dimension (see Theorem~\ref{thm:Gaussian} and the discussion both before and after that theorem).  We conclude the paper with a few brief remarks and suggestions.

\section{Review of ground states, critical droplets, and metastates}
\label{sec:review}

In this section we define the relevant quantities for our study and review results obtained in previous work.  For a more comprehensive treatment, we refer the reader to~\cite{NS22}.

\subsection{Ground states}
\label{subsec:gs}

The Edwards-Anderson (EA) Ising spin glass model~\cite{EA75} in zero magnetic field on the $d$-dimensional cubic lattice $\mathbb{Z}^d$ is defined by the Hamiltonian
\begin{equation}
\label{eq:EA}
{\cal H}_J=-\sum_{<x,y>} J_{xy} \sigma_x\sigma_y 
\end{equation}
where $\sigma_x=\pm 1$ is the Ising spin at site $x$ and $\langle x,y\rangle$ denotes an edge (or ``bond'' --- we will use the two terms interchangeably) in the nearest-neighbor edge set $\mathbb{E}^d$. Each edge $\langle x,y\rangle\in\mathbb{E}^d$ is assigned a coupling $J_{xy}$. The $J_{xy}$'s are independent, identically distributed continuous random variables chosen from a distribution $\nu(dJ_{xy})$.  Our requirements on $\nu$ are that it be distributed symmetrically about zero and that its moments are all finite; throughout most of what follows we will use a Gaussian distribution with mean zero and variance one.  We denote by $J$ a particular realization of the couplings.

Our focus is on ground states of the EA~spin glass in finite dimensions $d\ge 2$. Define $\Lambda_L$ to be a cube of side~$L$ centered at the origin; then a finite-volume ground state~$\sigma_L$ is the lowest-energy spin configuration in~$\Lambda_L$ subject to a specified boundary condition. An infinite-volume ground state~$\sigma$ can be defined in two equivalent ways: first, as any convergent $L\to\infty$ limit of a sequence of $\sigma_L$'s, or second, as a spin configuration $\sigma$ on all of $\mathbb{Z}^d$ defined by the condition that its energy cannot be lowered by flipping any {\it finite\/} subset of spins.  The condition for $\sigma$ to be a ground state is then that
\begin{equation}
\label{eq:gs}
E_{S}\equiv\sum_{\langle x,y\rangle\in{S}}J_{xy}\sigma_x\sigma_y\ >0 
\end{equation}
where $S$ is any closed surface (or contour in two dimensions) in the dual lattice. (We have abused notation somewhat by writing $\langle x,y\rangle\in S$ in the sum. This should be understood as meaning, ``sum over edges in the original lattice whose duals belong to $S$.'')  The surface~$S$ encloses a connected set of spins (a ``droplet''), and $\langle x,y\rangle\in{S}$ is the set of edges connecting spins inside~$S$ to spins outside~$S$.   The inequality in~(\ref{eq:gs}) is strict since, by the continuity of $\nu(dJ_{xy})$, there is zero probability of any closed surface having exactly zero energy in $\sigma$. Of course the condition~(\ref{eq:gs}) must also hold for finite-volume ground states $\sigma_L$ for any $S$ completely inside $\Lambda_L$.\footnote{Some authors (see, e.g., \cite{vEvH84}) use the term ``ground state'' to refer to a probability measure supported on ground state configurations.}

Given the spin-flip symmetry of the Hamiltonian, a ground state, whether finite- or infinite-volume, generated by a spin-symmetric boundary condition, such as free or periodic, will appear as one part of a globally spin-reversed pair; we therefore refer generally to ground state pairs (GSP's) rather than individual ground states, and denote both by $\sigma$ when the context is clear.  Clearly $\sigma$ must be defined with respect to a specific~$J$, but we suppress its dependence on $J$ for notational convenience.

\subsection{Critical droplets and flexibility}
\label{subsec:cd}

We turn next to critical droplets, which were introduced in~\cite{NS2D00,NS2D01} and whose properties were described extensively in~\cite{NS22} (see also~\cite{ANS19,ANS21,Nussinov24}).  Again we summarize only those properties relevant to the current study.  We begin with definitions (all of which should be understood as pertaining to some fixed coupling realization~$J$, which will generally be dropped for notational convenience). We begin with a heuristic discussion to motivate the definitions that follow.

For fixed coupling realization $J$, consider a finite-volume ground state $\sigma_L^>$ and a specific bond $b_i$ with coupling value $J({b_i})$. Suppose $J(b_i)=K$ in $J$ and that it is satisfied in $\sigma_L^>$; for the purpose of this discussion we take $K>0$. We will allow $J(b_i)$ to vary with all other couplings held fixed. As $J(b_i)$ increases above $K$, $\sigma_L^>$ becomes more stable and its spin configuration is unchanged. It will also remain unchanged (though with decreasing stability) for some finite range of values of $J(b_i)$ below $K$. Eventually, below some (positive or negative) value $J(b_i)<K$, the ground state becomes unstable and a droplet (a connected set of spins) overturns, leading to a new ground state $\sigma_L^<$.  We denote the critical value of $J(b_i)$ which separates $\sigma_L^>$ from $\sigma_L^<$ as $J_c^{\sigma_L} (b_i)$ (a formal definition appears below). It is easy to see that decreasing $J(b_i)$ further below $J_c^{\sigma_L}(b_i)$ now {\it increases\/} the stability of $\sigma_L^<$. 

The conclusion is that varying $J(b_i)$ from $+\infty$ to $-\infty$ while holding all other couplings fixed leads to a pair of ground states $\sigma_L^>$ and $\sigma_L^<$ (not the same as a GSP which refers to ground states which are global flips of each other at fixed~$J$), differing by a droplet flip. Specifically, there is a critical value $J_c^{\sigma_L}$, determined by all couplings {\it except\/} $J(b_i)$, such that for $J(b_i)>J_c^{\sigma_L}$, the ground state is $\sigma_L^>$, while for $J(b_i)<J_c^{\sigma_L}$, the ground state is $\sigma_L^<$.

What happens exactly at $J_c^{\sigma_L}$? It is not hard to see that precisely at that value, in both $\sigma_L^>$ and $\sigma_L^<$, there is a droplet of spins enclosed by a (shared) unique surface ${S}_i$ in the dual lattice which includes the dual edge $b_i^*$ and has precisely zero energy $E_{S_i}$ as defined in~(\ref{eq:gs}), with every other surface in the dual lattice having strictly positive energy.  The violation of~(\ref{eq:gs}) is allowed because at $J_c^{\sigma_L}$ the coupling $J(b_i)$ is not independent of the others; it has been tuned to infinite precision, with its value determined by the other couplings in~$\mathbb{E}_L^d$ (the set of edges whose endpoints are contained in $\Lambda_L$).

With this in mind, we make the following definitions.

\medskip

{\bf Definition [Newman-Stein~{\cite{NS2D00}]}}.  Consider the finite-volume GSP $\sigma_L$ for the EA Hamiltonian~(\ref{eq:EA}).  Choose a bond $b_i$ and consider all surfaces in the dual edge lattice ${\mathbb E}_L^*$ which include the dual edge $b_i^*$ and which partition the spins in $\Lambda_L$ into two disjoint sets. The energies of these surfaces are given by~Eq.~(\ref{eq:gs}) and so are all positive.  Because (by continuity of the coupling distribution) there is zero probability that any two such surfaces have equal energy, there must exist one of {\it least\/} energy in $\sigma_L$. We call this surface the {\it critical droplet boundary\/} of $b_i$ in $\sigma_L$ and denote it by $\partial D(b_i,\sigma_L)$. We further define the {\it critical droplet\/} of $b_i$ in $\sigma_L$ as the set of spins $D(b_i,\sigma_L)$ enclosed by $\partial D(b_i,\sigma_L)$.

\medskip

{\bf Remark.} The definition of critical droplets is not restricted to closed surfaces entirely within $\Lambda_L$; i.e., it is possible for a critical droplet to reach the boundary $\partial\Lambda_L$, with the proviso that the droplet, if overturned, must still obey the imposed boundary conditions. Hence a critical droplet reaching the boundary is ruled out for fixed boundary conditions but is allowed for free, periodic, or antiperiodic boundary conditions. In the case of free boundary conditions, a critical droplet reaching the boundary will not be a closed surface within $\Lambda_L$ (excluding $\partial\Lambda_L$); if it touches two separate faces of $\partial\Lambda_L$ it would then divide the spins in $\Lambda_L$ into two disjoint components both of which extend to the boundary. For periodic boundary conditions, the critical droplet boundary is a closed surface enclosing a connected droplet of spins in the equivalent $d$-dimensional torus, but the surface may not appear closed when viewed within the cube $\Lambda_L$.

\medskip

A few further remarks on terminology and notation: Critical droplets are defined with respect to edges rather than associated couplings to avoid confusion, given that we often vary the coupling value associated with specific edges, while the edges themselves are fixed, geometric objects.  The critical droplet $D(b_i,\sigma_L)$ and its boundary $\partial D(b_i,\sigma_L)$ are geometrically the same in both $\sigma_L^>$ and $\sigma_L^<$, so we simply use $\sigma_L$ to refer to the GSP under discussion; similarly for $J_c^{\sigma_L}$.  $\sigma_L^>$ and $\sigma_L^<$ differ by a rigid flip of the spins contained in $D(b_i,\sigma_L)$, so couplings in $\partial D(b_i,\sigma_L)$ which are satisfied in $\sigma_L^>$ are unsatisfied in $\sigma_L^<$ while those that are unsatisfied in $\sigma_L^>$ are satisfied in $\sigma_L^<$.  No other couplings in $\sigma_L$ change their satisfaction status as $J(b_i)$ is varied from $\infty$ to $-\infty$.
\medskip

We next  define the energy~$E\bigl(D(b_i,\sigma_L)\bigr)$ of the critical droplet of $b_i$ in $\sigma_L$ to be the energy of its boundary as given by~(\ref{eq:gs}):
\begin{equation}
\label{eq:cd}
E\bigl(D(b_i,\sigma_L)\bigr)=\sum_{<x,y>\in\partial D(b_i,\sigma_L)}J_{xy}\sigma_x\sigma_y\, .
\end{equation}

\medskip

{\bf Definition [Newman-Stein~{\cite{NS2D00}]}}. The {\it critical value\/} in $\sigma_L$ of the coupling $J(b_i)$ is denoted $J_c^{\sigma_L}(b_i)$ (or simply $J_c^{\sigma_L}$ if the bond in question is unambiguous) and is the value of $J(b_i)$ where $E\bigl(D(b_i,{\sigma_L})\bigr)=0$, while all other couplings in $J$ are held fixed.

\medskip 

{\bf Definition [Newman-Stein~{\cite{NS2D00}]}}. Let $J_c^{\sigma_L}(b_i)$ be the critical value of $J(b_i)$ in $\sigma_L$.  Suppose $J(b_i)=K$ in $J$. We define the {\it flexibility\/}~of $J(b_i)$ at that particular value to be $f(J(b_i),{\sigma_L})=\vert K-J_c^{\sigma_L}(b_i)\vert$. 

\medskip

{\bf Remark.} The critical value $J_c^{\sigma_L}(b_i)$ is determined by all couplings in $J$ {\it except\/} $J(b_i)$.  Because couplings are chosen independently from $\nu(dJ_{xy})$, it follows that the value $J(b_i)$ is {\it independent\/} of~$J_c^{\sigma_L}(b_i)$. Therefore, given the continuity of $\nu(dJ_{xy})$, for arbitrary~$J$ there is zero probability in a ground state that any coupling has exactly zero flexibility; for an arbitrary coupling realization~$J$ all flexibilities are strictly positive with probability~one.

\smallskip

It follows from the definitions above that
\begin{equation}
\label{eq:flex}
f(J(b_i),\sigma_L)=E\bigl(D(b_i,\sigma_L)\bigr)\, .
\end{equation}
Therefore couplings which share the same critical droplet have the same flexibility. 

All of the above definitions work equally well whether the GSP under discussion is finite- or infinite-volume.  We note that a complete analysis of critical droplets and flexibilities within infinite-volume ground states requires use of the excitation metastate, whose definition and properties were presented in~\cite{NS2D00,NS2D01,ADNS10,ANS19}, to which we refer the interested reader.  The important conclusion from those studies is that finite-volume critical droplets and their associated flexibilities converge with their properties preserved in the infinite-volume limit (cf.~Lemma~3.1 from~\cite{NS22}). This is true even for a critical droplet that in the limit is infinite in extent in one or more directions (if such exist).  Excitation metastates can then be used to define unbounded critical droplets which enclose an infinite subset of spins: they are the infinite-volume limits of critical droplets in finite-volume ground states.  Of particular importance is that infinite-volume droplets possess a well-defined energy in the infinite-volume limit~\cite{NS2D01,ADNS10,ANS19,NS22}.\footnote{However, one cannot rule out the possibility that such a critical droplet might become disconnected in the infinite-volume limit. If so, this does not affect our results: as we will see in Sect.~\ref{sec:sfcd}, our arguments rely on a single coupling ``controlling the flexibility'' of a positive density of other couplings in certain critical droplets. That this property is retained in the infinite-volume limit regardless of details of the critical droplet geometry can be seen by the reversibility of the process of flipping a critical droplet: if changing the coupling value~$J(b_0)$ of a particular bond~$b_0$ from $J_1$ to $J_2$ flips the critical droplet of $J(b_0)$, then in reversing the process from $J_2$ to $J_1$ the original GSP is restored.  This means that every coupling that changed from satisfied to unsatisfied or vice-versa in changing $J(b_0)$ from $J_1$ to $J_2$  must change back to its original satisfaction status on changing~$J(b_0)$ from $J_2$ to $J_1$. This can happen only if $J(b_0)$ controls the flexibility of every coupling whose satisfaction status changes when passing through the critical value of $J(b_0)$ in either direction.}

How might such unbounded critical droplets arise? A natural construction is to consider a sequence of volumes $\Lambda_L$ with the corresponding $\sigma_L$'s converging to an infinite-volume ground state (or GSP) $\sigma$.  Suppose there exists an edge $b_i$ whose finite-volume ground state critical droplet boundaries, though finite in every $\sigma_L$, increase in size without bound as $L\to\infty$ with their corresponding flexibilities monotonically decreasing as $L$ increases. In the limit $L\to\infty$ one then arrives at a critical droplet with infinite boundary comprising an infinite subset (with respect to $\mathbb{Z}^d$) of spins in $\sigma$ and with a well-defined (and still strictly positive) limiting energy. 

{\bf Remark.}  We noted above that as a coupling $J(b_i)$ passes through its critical value $J_c^\sigma$, say from $J_c^\sigma+\epsilon$ to $J_c^\sigma-\epsilon$, the ground state changes from $\sigma^>$ to $\sigma^<$ due to the flip of the critical droplet $D(b_i,\sigma)$.  In the case where $D(b_i,\sigma)$ is infinite, a question arises:  could it happen that when $J(b_i)=J_c^\sigma-\epsilon$, $\sigma^>$ retains the property~(\ref{eq:gs}) and therefore remains a ground state coexisting with $\sigma^<$?  This can only occur --- if it 
\noindent occurs at all --- for a limited range of values of $J(b_i)$: when $J(b_i)<J_c^\sigma$ and $\vert J(b_i)\vert>J^{\rm ub{}}$ (see~Eq.~(\ref{eq:ss})), $\sigma^>$ can no longer satisfy~(\ref{eq:gs}).  Although we can't rigorously rule out the possibility that $\sigma^>$ retains its ground state property~(\ref{eq:gs}) for some finite range of coupling values $J(b_i)<J_c^\sigma$, heuristically it seems unlikely. For example, if an infinite critical droplet arises as suggested in the previous paragraph, then when $J(b_i)=J_c^\sigma-\epsilon$, for any $\epsilon>0$, there will be an {\it infinite\/} sequence of finite critical droplets violating~(\ref{eq:gs}).  

 \vfill\eject
 
\subsection{Metastates}
\label{subsec:meta}

The concept of the metastate has been introduced and discussed in multiple papers~\cite{AW90,NS96c,NSBerlin,NS97,NS98,NS2D00,NS2D01,NS03b,Read14,ADNS10,ANSW14,ANSW16,ANS19,Read22, NRS23b,NS24a,NS24b, Read24}, and provides a setting for working with infinite-volume spin glasses at zero or positive temperature.  
For details, we refer the reader to those papers; in particular, Ref.~\cite{NRS23b} contains a comprehensive discussion.  In this paper we focus on the zero-temperature metastate.

A metastate is a probability measure on infinite-volume Gibbs states.  Suppose one examines an infinite sequence of volumes $\Lambda_L$ each with a specified boundary condition.  Depending on the Hamiltonian, temperature, and boundary conditions chosen, this sequence of finite-volume Gibbs states might converge to a single (pure or mixed) infinite-volume Gibbs state (i.e., a thermodynamic state), or else it may not converge but have two or more subsequences converging to different Gibbs states. Informally, the metastate is a probability measure that describes the distribution of these distinct thermodynamic states, or equivalently, it describes the distribution of the collection of all correlation functions within a large arbitrary volume. 

Formally, a metastate is a probability measure on infinite-volume Gibbs states, depending on $J$ and inverse temperature~$\beta$, and satisfying the properties of {\it coupling-\/} and {\it translation-covariance\/}.  The latter simply requires that a uniform lattice shift does not affect the metastate properties. This can be expressed as the following requirement: for any lattice translation $\tau$ of $\mathbb{Z}^d$ and a subset $A$ of probability measures on the space of spin configurations $\{-1,+1\}^{\mathbb{Z}^d}$,
\begin{equation}
\label{eq:trans}
\kappa_{\tau J}(A)=\kappa_{J}(\tau^{-1}A).
\end{equation}
This is guaranteed when one constructs a metastate using periodic boundary conditions to generate the finite-volume Gibbs states; in the infinite-volume limit, the Gibbs states (and therefore the metastate) will inherit the torus-translation covariance of the finite-volume Gibbs states.  However, one can also construct translation-covariant metastates using fixed or free boundary conditions by taking translates in a prescribed manner~\cite{NS2D01}.


Coupling covariance refers to transformations on states under finite changes in the values of a finite number of couplings.  Changing a finite set of couplings will change the thermodynamic states, i.e., the correlation functions.  However, it was shown that under a finite change of couplings, a pure state transforms to a pure state~\cite{AW90,NS96c}, and therefore a convex mixture of multiple pure states (i.e., a mixed Gibbs state) remains a convex mixture of the transformed pure states, generally with modified weights.  Coupling covariance can be expressed as follows:  for  $B$ a finite subset of $\mathbb{Z}^d$, $J_B$ the set of couplings assigned to the edges in $B$, $f(\sigma)$ a function of a finite set of spins, and $\Gamma$ a Gibbs state, we define the operation ${\cal L}_{J_B} : \Gamma \mapsto {\cal L}_{J_B}\Gamma$ by its effect on the expectation $\langle\cdots\rangle_\Gamma$ 
in $\Gamma$:
\begin{equation}
\label{eq:gamma L}
 \left\langle f(\sigma)\right\rangle_{{\cal L}_{J_B}\Gamma}= \frac{\left\langle f(\sigma) \exp\Bigl(-\beta H_{J_B}(\sigma)\Bigr)\right\rangle_\Gamma}
{\left\langle\exp\Bigl(-\beta H_{J_B}(\sigma)\Bigr)\right\rangle_\Gamma}\, ,
\end{equation}
which describes the effect of modifying the couplings within $B$. We require that the metastate 
be covariant under local modifications of the couplings, i.e., for any subset $A$ defined as in~(\ref{eq:trans}),
\begin{equation}
\label{eq:cc}
\kappa_{J+J_B}(A)= \kappa_{J}({\cal L}_{J_B}^{-1}A)\, ,
\end{equation}
where ${\cal L}_{J_B}^{-1} A$ equals the set of $\Gamma$'s such that ${\cal L}_{J_B}\Gamma\in A$.  (At zero temperature, the treatment of coupling covariance is best done in the setting of the excitation metastate; see~\cite{ANSW16} for details.)
In other words, the set of Gibbs states on which the metastate is supported does not change, aside from the usual changes in correlation functions within the individual states.
In particular, no Gibbs states either flow into or out of the metastate under a finite change of couplings.  

The metastate of interest in this paper is a {\it zero-temperature periodic boundary condition metastate\/}, denoted $\kappa_J(\sigma)$ (or often simply $\kappa_J$), which is a probability measure on infinite-volume ground state pairs~$\sigma$ induced by an infinite sequence of volumes with periodic boundary conditions using the EA~Hamiltonian~(\ref{eq:EA}); it is the marginal distribution of the excitation metastate.  We will refer to the more general class of zero-temperature translation-covariant EA metastates (of which $\kappa_J$ is a member) by ${\cal N}_J$, and we will denote a generic member of  ${\cal N}_J$ by $\eta_J$.

\subsection{Properties of critical droplets}
\label{subsec:properties}

In this section we review some earlier results which will be needed in what follows. Proofs will mostly be omitted; we refer the interested reader to the references where they appear.  From here on we work exclusively with infinite-volume GSP's denoted by $\sigma$.

\medskip

\begin{lemma} 
\label{lem:share}
{\rm (Newman-Stein~\cite{NS22}, Lemma~2.5).  Consider two distinct edges $b_1$ and $b_2$ and an infinite-volume ground state $\sigma$. (a) If $f(J(b_1),\sigma) > f(J(b_2),\sigma)$, then $b_1$ cannot belong to $\partial D(b_2,\sigma)$, while $b_2$ may or may not belong to~${\partial D}(b_1,\sigma)$.  (b) If $b_1$ and $b_2$ share the same critical droplet, then w.p.~1 $b_1\in\partial D(b_2,\sigma)$ and $b_2\in\partial D(b_1,\sigma)$ (the converse is true as well).  If $b_1$ and $b_2$ share the same critical droplet, then by Eq.~(\ref{eq:flex}) $J(b_1)$ and $J(b_2)$ have equal flexibilities.}
\end{lemma}

\medskip

\begin{lemma} 
\label{lem:remain}
{\rm (Newman-Stein~\cite{NS22}, Lemma~6.2). Suppose a bond $b_1$ with coupling value $J_1$ in $J$ and critical value $J^\sigma_c$ in $\sigma$ belongs to the critical droplet boundary $\partial D(b_2,\sigma)$ of a different bond $b_2$. Then $b_1$ will remain in $\partial D(b_2,\sigma)$ for the entire range of coupling values between $J_1$ and $J^\sigma_c$.}
\end{lemma}

\medskip

\begin{lemma} 
\label{lem:flexlower}
{\rm If the flexibility of any given coupling is lowered (by changing its coupling value) but remains positive in $\sigma$, the flexibility of any other edge in $\sigma$ is either also lowered (by up to the same amount) or else remains unchanged.  Similarly, if the flexibility of any coupling is increased, then the flexibility of any other edge in $\sigma$ is either also raised (by up to the same amount) or else remains unchanged.}
\end{lemma}

\smallskip

{\bf Remark.} This is an extension of Lemma~2.6 of~\cite{NS22}. There was an error in the statement of that lemma, which claimed that lowering the flexibility of a coupling either lowered the flexibility of other couplings by the same amount (instead of up to the same amount) or else left the flexibility unchanged.  That had no effect on any of the subsequent conclusions of the paper, but we take this opportunity to correct it. Lemma 2.6 in~\cite{NS22} did not discuss raising the flexibility.

\medskip

{\bf Proof of Lemma~\ref{lem:flexlower}.}  Choose an arbitrary bond $b_0$ with running coupling value $J(b_0)$, and suppose $J(b_0)=J_0$ in $J$ and has critical value $J^\sigma_c<J_0$ in $\sigma$.  Changing the initial coupling value from $J_0$ to a lower value $J_1$ with $J_1\in(J^\sigma_c,J_0)$ lowers the flexibility of $J(b_0)$ without affecting $\sigma$. The question then becomes whether it affects the flexibilities of other couplings in $\sigma$. 

There are four types of bonds to consider. The first, which we call Type~1 bonds, are those which share the same critical droplet $D(b_0,\sigma)$ when $J(b_0)=J_0$; by Lemma~\ref{lem:share} all Type~1 bonds lie in~${\partial D}(b_0,\sigma)$. Type~2 bonds are those which do not lie in ${\partial D}(b_0,\sigma)$ but whose critical droplet boundaries include $b_0$.  Type~3 are bonds which belong to~${\partial D}(b_0,\sigma)$ but whose critical droplets are other than $D(b_0,\sigma)$ when $J(b_0)=J_0$. Type~4 are all other bonds.

Consider first Type~1 bonds which share the critical droplet $D(b_0,\sigma)$. By Eq.~(\ref{eq:flex}) all such bonds have the same flexibility as $J(b_0)$, so their flexibility is lowered by the same amount as that of $J(b_0)$.   

Similarly, the critical droplet boundaries of Type~2 bonds include~$b_0$ though they themselves do not lie in~${\partial D}(b_0,\sigma)$ when $J(b_0)=J_0$. By Lemma~\ref{lem:share}, the flexibility of a Type~2 bond is greater than that of $J_0$, so when $J(b_0)$ is lowered without passing through its critical value, the flexibility of a Type~2 bond is lowered by the same amount with no droplet flip occurring.

Although Type~3 bonds belong to ${\partial D}(b_0,\sigma)$, their critical droplets have energies less than $E(D(b_0,\sigma))$ when $J(b_0)=J_0$, so at first their flexibilities will remain unchanged. As $J(b_0)$ approaches $J_c^+$, $E(D(b_0,\sigma))$ will become less than the critical droplet of any Type~3 bond, so if $J_1=J_c^+$ the critical droplet of any Type~3 bond changes to $D(b_0,\sigma)$ at some $J(b_0)\in(J_c^+,J_0)$;  below that value its flexibility decreases. Therefore, over the entire process its flexibility decreases by an amount smaller than that of $J(b_0)$. 

Type~4 bonds are those whose critical droplet boundaries remain disjoint from ${\partial D}(b_0,\sigma)$ as $J(b_0)$ changes from $J_0$ to $J_1$, and so their flexibilities remain unchanged. This proves the first part of the lemma.

The second part of the lemma concerns starting $J(b_0)$ at a fixed value $J_0$ and then moving the coupling value {\it away\/} from~$J_c^\sigma$; e.g., if the starting value of $J(b_0)=J_0>J_c^\sigma$ then  its final value is $J_2>J_0$. Now the flexibilities of type~1 bonds that remain in ${\partial D}(b_0,\sigma)$ throughout the entire process will increase by the maximum amount $\Delta J=J_2-J_0$.
However, there may be other bonds $b_i\in {\partial D}(b_0,\sigma)$ which initially remain in ${\partial D}(b_0,\sigma)$, but as $J(b_0)$ continues to increase, will switch at some $J(b_0)$ to a different critical droplet and will remain in that new droplet as $J(b_0)$ continues to increase. Their final energy change will increase by an amount strictly smaller than $\Delta J$. The same argument and conclusion applies to Type~2 bonds.

Because Type~3 bonds already belong to droplets with lower energy than $E(D(b_0,\sigma))$ when $J(b_0)=J_0$, they will remain in those critical droplets as $J(b_0)$ moves away from $J_c^\sigma$, and similarly for Type~4 bonds, so the energies of both types of bonds remains unchanged. This completes the proof of the second part of the theorem.~$\diamond$

\medskip

A useful quantity in what follows~\cite{NS22} is the empirical probability distribution $P_J(f,\sigma)$ over all edges of the flexibilities $f$ in the ground state $\sigma$, constructed by computing the finite-volume flexibility distribution in an increasing sequence of volumes and taking successive averages (with equal weight on each finite volume, as in a microcanonical ensemble approach) of this distribution. Convergence in the infinite-volume limit (for a.e.~$\sigma$ in the support of the metastate $\eta_J$ for a.e.~$J$) follows from the ergodic theorem.  In this paper we will use only the mean $\langle f\rangle_\sigma$ over all edges of the flexibilities $f$ in the ground state $\sigma$. As will be seen following the statement of Theorem~\ref{thm:flexconst} below, given our conditions on $\nu(dJ_{xy})$ this is a finite quantity for a.e.~$J$.  

\medskip

\begin{theorem}
\label{thm:flexconst}
{\rm (adapted from Newman-Stein~\cite{NS22}, Lemma~7.2).  Let $\langle f\rangle_J$ be the metastate average of $\langle f\rangle_\sigma$ over the ground states~$\sigma$ in the support of the metastate~$\eta_J$.  Then $\langle f\rangle_J=\langle f\rangle$ is almost surely~constant (i.e., constant except for a set of measure zero) with respect to $J$. }
\end{theorem}

Although in this paper we confine ourselves to the mean of the flexibility over all edges, it is not difficult to show that all moments of the flexibility are finite for a.e.~$J$ if one uses a distribution $\nu(dJ_{xy})$ on the couplings satisfying the requirements listed in the discussion following~(\ref{eq:EA}). We omit a detailed proof here but provide a brief sketch of an argument for the mean; finiteness of higher moments can be shown using a similar argument.   For specificity we will use a Gaussian distribution of mean zero and variance~one for the couplings.   Although the argument to follow will work for spins 
and couplings on any bipartite graph, we confine the discussion as before to $\mathbb{Z}^d$.  

For a fixed coupling realization~$J$ and a GSP consistent with it, consider all of the spins on the sites in one of the two sublattices of $\mathbb{Z}^d$. 
Each of these spins lies at an endpoint of $2d$ i.i.d.~couplings; we will say these couplings are ``connected to'' the given spin. 
No two sublattice spins share a coupling, and the union of the couplings connected to all of the sublattice spins constitute all of the couplings in the edge set $\mathbb{E}^d$.   
In what follows, then, the spins will always refer only to those on one sublattice (hence they comprise half the spins in the entire system) while the couplings 
refer to all of the couplings assigned to the edges in $\mathbb{E}^d$.

Next choose a particular spin $\sigma_x$.  If we overturn $\sigma_x$, we get a (one-spin) droplet whose
boundary includes all of the $2d$ edges $b_{xz}$ with $\vert z-x\vert=1$.  The sum of the absolute values 
of each of the $2d$ couplings $J_{xz}$ is then an upper bound to the flexibility of each of these couplings, because by~(\ref{eq:flex}) the flexibility of a coupling equals the energy
of its critical droplet, and the actual critical droplet of a coupling is the lowest-energy droplet whose boundary includes that coupling.

So far we have found that an upper bound for the flexibility of each of the $2d$ edges connected to $\sigma_x$ is the sum of the absolute values of the 
$2d$ i.i.d.~Gaussian couplings connected to $\sigma_x$. Call this upper bound $B_x$; because the couplings are all independent
their means and variances add, so the mean of $B_x$ is $2d\sqrt{2/\pi}$ and its variance is $2d(1-2/\pi)$.

Furthermore, because the couplings connected to $\sigma_x$ are independent of the couplings connected to $\sigma_y$ 
for all $y\ne x$, {\it every\/} coupling $\sigma_y$ in the lattice has an upper bound $B_y$ on its flexibility arrived at in the same way.
The set of all $\{B_x:\ x\in\mathbb{Z}^d\}$ is then a collection of i.i.d.~random variables with mean and variance given above.
By ergodicity, the mean of $B_x$ over all sites $x\in\mathbb{Z}^d$ is equal to the mean of the distribution of a single $B_x$, 
which as before is~$2d\sqrt{2/\pi}$.

By definition all flexibilities are strictly positive, so $2d\sqrt{2/\pi}$~is an upper bound of the mean of the flexibilities of all couplings for a.e.~$J$.
A similar argument using non-Gaussian coupling distributions~$\nu(dJ_{xy})$ can be used for the mean and higher moments of the flexibility as long as the corresponding moments 
of $\nu(dJ_{xy})$ are finite.

\subsection{Types of critical droplets}
\label{subsec:types}

The nature of critical droplets in a one-dimensional spin glass is trivial: for every bond $b_i$ in the system, its critical droplet boundary consists of $b_i$ only, and its critical droplet consists of a semi-infinite set of spins~\cite{NS22}.  From here on, we confine ourselves to dimensions $d\ge 2$.


Critical droplets in $d\ge 2$ can be bounded, enclosing a finite set of spins, or infinite in extent, separating the spins in ${\mathbb Z}^d$ into two infinite disjoint subsets. Our main concern in what follows is not the droplet $D$ itself (i.e., the spins which flip as a coupling passes through its critical value)  but rather its boundary ${\partial D}$ (i.e., the set of edges separating the region of flipped spins from that of unflipped spins when a coupling passes through its critical value).   From this perspective there are three kinds of critical droplets: those whose boundaries are finite, those whose boundaries consist of an infinite set of edges with zero density in ${\mathbb E}^d$ (these typically have $d_s<d$, where $d_s$ is the dimension of the boundary and $d$ the space dimension), and those where $d_s=d$ and whose boundaries consist of an infinite set of edges with positive upper density (from here on, we will simply refer to `positive density', which should be understood as positive upper density). The latter are of particular importance.

\medskip

{\bf Definition}. Consider an edge $b_{xy}$ and an infinite-volume ground state $\sigma$. We will say that  ``the critical droplet of $b_{xy}$ in $\sigma$ is space-filling'' to mean that $\partial D({b_{xy},\sigma})$ comprises a positive density of bonds in $\mathbb{E}^d$.

\medskip

We will hereafter refer to a droplet whose boundary comprises a positive density of bonds in $\mathbb{E}^d$ as a {\it space-filling critical droplet\/} (SFCD).  (Note that ``space-filling'' is used here in a different way that in the classical setting of curves in Euclidean space.)  We refer to a critical droplet whose boundary is infinite but comprises a zero density of bonds as a {\it zero-density\/} critical droplet. The third kind of critical droplet is bounded in space and encloses a finite set of spins; this will be referred to as a finite critical droplet.

\medskip

\begin{theorem}
\label{thm:three}
{\rm (Newman-Stein~\cite{NS24c}, Theorem~3.1).  Let $\sigma$ denote an infinite-volume spin configuration.  Then for almost every $(J,\sigma)$ pair at zero temperature (which restricts the set of $\sigma$'s to ground states corresponding to particular coupling realizations~$J$), and for any of the three types of critical droplet (finite, zero-density, or positive-density), either a positive density of edges in $\sigma$ has a critical droplet of that type or else no edges do.}
\end{theorem}

\section{Continuity of flexibility}
\label{sec:continuity}

In Sect.~\ref{sec:sfcd} we will study how the quantity $P_J(f)$ (defined above Theorem~\ref{thm:flexconst}) behaves as the value of an arbitrary single coupling $J(b_0)$ is varied with all others held constant.  Before doing so we need to establish how the flexibility in a GSP $\sigma$ is affected when $J(b_0)$ passes through $J_c^\sigma$ in either direction.

We begin by looking at the behavior of the flexibility $f(J(b_0),\sigma)$ of $J(b_0)$ itself. By the definition of flexibility, when $J(b_0)=J_0$ its flexibility is $f(J(b_0),\sigma)=\vert J_0-J_c^\sigma(b_0)\vert$.  Because $J_c^\sigma(b_0)$ depends on all other couplings in $J$ {\it except\/} $J(b_0)$, it follows immediately that if $J(b_0)$ is varied while holding all other couplings fixed, $f(J(b_0),\sigma)$ varies continuously with $J(b_0)$, including when $J(b_0)$ passes through $J_c^\sigma(b_0)$.  

Moreover, by Eq.~(\ref{eq:flex}) the flexibility $f(J(b_0),\sigma)$ equals the energy of the critical droplet $D(b_0,\sigma)$, i.e., $f(J(b_0),\sigma)=E(D(b_0,\sigma))$.  Suppose the critical droplet of a different bond $b_i$ is also $D(b_0,\sigma)$, i.e., $b_0$ and $b_i$ share the same critical droplet in $\sigma$.  By Lemma~\ref{lem:share} this can only occur for bonds $b_i\in{\partial D}(b_0,\sigma)$, the boundary of $D(b_0,\sigma)$.  As a consequence, the flexibility of $J(b_i)$ will also equal $\vert J(b_0)-J_c^\sigma\vert$.  As noted, all couplings $b_i$ that share the critical droplet $D(b_0,\sigma)$ (and hence have the same flexibility) are in ${\partial D}(b_0,\sigma)$, but the converse is not necessarily true unless $J(b_0)$ is sufficiently close to $J_c^\sigma(b_0)$ (Theorem 6.3 of~\cite{NS22}).  That is, when $J(b_0)$ is far from its critical value, not all couplings $b_i\in{\partial D}(b_0,\sigma)$ may have $D(b_0,\sigma)$ as their critical droplet, but as shown in Theorem~6.3 of~\cite{NS22}, when $J(b_0)$ is sufficiently close to $J_c^\sigma(b_0)$, {\it all\/} couplings $b_i\in{\partial D}(b_0,\sigma)$ share the critical droplet $D(b_0,\sigma)$. Therefore, using similar reasoning as in the proof of Lemma~\ref{lem:flexlower}, if $J(b_0)$ changes by an amount $\Delta J(b_0)$ (again regardless of whether or not $J(b_0)$ passes through $J_c^\sigma(b_0)$) the change in flexibility of every coupling in ${\partial D}(b_0,\sigma)$ is less than or equal to $\Delta J(b_0)$.

This establishes that for all bonds $b_i\in{\partial D}(b_0,\sigma)$, when $J(b_0)$ is changed by $\Delta J(b_0)$ the flexibilities $f(J(b_i))$ can change by no more than $\Delta J(b_0)$, regardless of the starting value of $J(b_0)$ or the size of $\Delta J(b_0)$.

Next consider a bond $b_j$ whose critical droplet and its boundary are disjoint from those of $b_0$, i.e. ${\partial D}(b_0,\sigma)\cap{\partial D}(b_j,\sigma)=\emptyset$.   All their flexibilities $f(J(b_j))$ remain constant as $J(b_0)$ varies, again irrespective of whether $J(b_0)$ passes through $J_c^\sigma(b_0)$.

\medskip

The remaining case is that of a bond, which we will refer to as $b_2$, that is not in the critical droplet of $b_0$ but whose critical droplet contains one or more bonds in ${\partial D}(b_0,\sigma)$ (see Fig.~2 in Appendix~A for an illustration of this situation).  In this case it is not {\it a priori\/} clear that the flexibility of $J(b_2)$ doesn't jump when $J_0$ passes through $J_c^\sigma(b_0)$: the critical droplet of $b_2$ contains bonds (in ${\partial D}(b_0,\sigma)$) which abruptly change from satisfied to unsatisfied, or vice-versa, when $J(b_0)$ passes through $J_c^\sigma(b_0)$.  We will demonstrate below (Corollary~\ref{cor:flexchange}), however, that there is no jump in the flexibility of $J(b_2)$ when $J(b_0)$ passes through $J_c^\sigma(b_0)$.

Consider a coupling realization $J$, a GSP $\sigma$, and a bond $b_2$ with coupling value $J(b_2)=J_2$ in $J$. Consider a second coupling realization $J'$ which is the same as $J$ except that now $J(b_2)$ is moved closer to its critical value $J_c^\sigma(b_2)$ without reaching it; call this new coupling value $J'_2$. $J$ and $J'$ are identical except for the coupling value associated with $b_2$; moreover, since the critical value $J_c^\sigma(b_2)$ hasn't been crossed, the GSP $\sigma$ is also the same in $J$ and $J'$. Finally, consider a separate bond $b_0$ with two properties: $\partial D(b_0,\sigma)$ shares at least one bond with $\partial D(b_2,\sigma)$ (as shown in Fig.~\ref{fig:1} in Appendix~A); and $\partial D(b_0,\sigma)$ {\it never\/} includes $b_2$ itself, in both $J$ and $J'$ regardless of the value of $J(b_0)$.  Because ${\partial D}(b_0,\sigma)$ never includes $b_2$, $J_c^\sigma(b_0)$ and $D(b_0,\sigma)$ (and of course $J_c^\sigma(b_2)$ and $D(b_2,\sigma)$) are all independent of $J(b_2)$ and so are unchanged in going between $J$ and $J'$.

\medskip

\begin{lemma}
\label{lem:same}
{\rm Consider the situation described above.  Vary $J(b_0)$ in both $J$ and $J'$ while holding all other couplings fixed.  The flexibility of (a different, fixed coupling) $J_2$ in $J$ is $\vert J_2-J_c^\sigma(b_2)\vert$ and that of $J'_2$ in $J'$ is $\vert J'_2-J_c^\sigma(b_2)\vert$.  Then as $J(b_0)$ varies, the {\it change\/} in flexibility of $J_2$ and of $J'_2$ will be the same, regardless of whether $J(b_0)$ passes through  $J_c^\sigma(b_0)$.}
\end{lemma}

\medskip

{\bf Proof.}  Under the conditions stated in the theorem, $D(b_0,\sigma)$ is the same in $J$ and $J'$, and similarly for $D(b_2,\sigma)$.  Moreover, the critical value $J_c^\sigma(b_0)$ is the same in $J$ and $J'$, and similarly for $J_c^\sigma(b_2)$. By definition the flexibility of $J_2$ in $\sigma$ is $f(J_2)=\vert J_2-J_c^\sigma(b_2)\vert$ and of $J'_2$ is $f(J'_2)=\vert J'_2-J_c^\sigma(b_2)\vert$.  Varying $J(b_0)$ can in principle change $D(b_2,\sigma)$ and its corresponding critical value $J_c^\sigma(b_2)$, but because both $D(b_2,\sigma)$ and $J_c^\sigma(b_2)$ are independent of $J(b_2)$, any change in $J_c^\sigma(b_2)$ must occur {\it simultaneously\/} (i.e., at the same value of $J(b_0)$) in $J$ and $J'$, so at any fixed value of $J(b_0)$,  $J_c^\sigma(b_2)$ is the same in both $J$ and $J'$.  Therefore any change in the  flexibility of $f(J_2)$ and $f(J'_2)$ must be identical. $\diamond$

\medskip


\begin{lemma}
\label{lem:outside}
{\rm Consider an arbitrary bond $b_0$ and a GSP $\sigma$ consistent with coupling realization $J$.  Consider a bond $b_2$ that is not in ${\partial D}(b_0,\sigma)$ for any value of $J(b_0)$, but whose critical droplet contains one or more bonds in ${\partial D}(b_0,\sigma)$ (see Fig.~\ref{fig:1} in Appendix~A). Then for any $\Delta J(b_0)$, and starting from {\it any} value $J_0$ of $J(b_0)$, lowering (raising) the coupling value to $J_0-\Delta J(b_0)$ ($J_0+\Delta J(b_0)$) while holding all other couplings fixed can change the flexibility of $J(b_2)$ by an amount no greater than $\Delta J(b_0)$.}
\end{lemma}

{\bf Remark.} There can be situations where a bond such as $b_2$ is not in ${\partial D}(b_0,\sigma)$ when $J(b_0)$ is far from its critical value, but becomes part of ${\partial D}(b_0,\sigma)$ when $J(b_0)$ approaches $J_c^\sigma(b_0)$ without reaching it (an example of this occurring can be found in Appendix~A).  In such situations, the discussion preceding the statement of Lemma~\ref{lem:same} already shows that the flexibility of $J(b_2)$ changes by an amount no greater than $\Delta J(b_0)$, so that case need not be separately considered. The only remaining case to consider then is when $b_2$ is not in ${\partial D}(b_0,\sigma)$ for any value of $J(b_0)$.

As before let $\sigma^>$ denote the ground state when $J(b_0)>J_c^\sigma(b_0)$ and $\sigma^<$ denote the ground state when $J(b_0)<J_c^\sigma(b_0)$.  Because $J_c^\sigma(b_0)$ and $D(b_0,\sigma)$ are the same in both $\sigma^>$ and $\sigma^<$, we hereafter write $D(b_0,\sigma)$ for the critical droplet of $b_0$ and $J_c(b_0)$ for its critical coupling value, unless a specification of the GSP is required.


\medskip

Next we prove Lemma~\ref{lem:outside}, and in Appendix~A we present a specific example to illustrate how it works in practice.

{\bf Proof of Lemma \ref{lem:outside}.}  We already know from Lemma~\ref{lem:flexlower} that the conclusions of Lemma~\ref{lem:outside} are valid for all bonds except possibly when $J(b_0)$ crosses its critical value, where {\it a priori\/} the flexibility could undergo a discontinuous jump of magnitude $\pm\Delta$ in some bonds as $J_c(b_0)$ is crossed. We wish to show that $\Delta=0$ for all bonds.

As already shown in the discussion preceding Lemma~\ref{lem:same}, the only bonds for which $\Delta$ could be nonzero are bonds not in the critical droplet of $b_0$ but whose critical droplet contains one or more bonds in ${\partial D}(b_0,\sigma)$. As before, we denote such a bond as $b_2$.  By Lemma~\ref{lem:same}, if there is a jump of magnitude $\vert\Delta\vert$ in the flexibility of $J(b_2)$ as $J(b_0)$ crosses $J_c(b_0)$, it will be the same for any value of $J(b_2)$ on one side of $J_c(b_2)$. Therefore, without loss of generality, we can take $J(b_2)$ to be sufficiently close to $J_c(b_2)$ so that $E(D(b_2))=0^+$.  Now let $J(b_0)$ move from just above $J_c(b_0)$ to just below. This will cause the critical droplet $D(b_0,\sigma)$ to flip, changing the ground state from $\sigma^>$ to $\sigma^<$ and with it the satisfaction status of all bonds in $\partial D(b_0,\sigma)$: satisfied couplings in $\sigma^>$ are unsatisfied in $\sigma^<$ and vice-versa.  By the definition of a critical droplet, these are the {\it only\/} couplings that change their satisfaction status when $J_c(b_0)$ is crossed.  If $\Delta<0$, when $J(b_0)$ crosses $J_c(b_0)$ from above to below the critical droplet energy of $b_2$ becomes $E(D(b_2)) = -\vert\Delta\vert$, which would also flip the critical droplet $D(b_2)$.  This will change the satisfaction status of $J(b_2)$, which cannot happen (only the couplings in ${\partial D}(b_0)$ will do so). Therefore $\Delta\ge 0$.

If one reverses the procedure, keeping everything fixed while changing $J(b_0)$ from $J_c^-$ back to $J_c^+$, the jump in flexibility of $J(b_2)$ must then be $-\vert\Delta\vert$.
But the previous argument demonstrates that the jump in flexibility of $J(b_2)$ cannot be negative when $J_c(b_0)$ is crossed in either direction, so $\Delta=0$. $\diamond$

\medskip

\begin{corollary}
\label{cor:flexchange}
{\rm In a GSP $\sigma$, if the flexibility of any edge is changed by an amount $\Delta J$, then the flexibility of any other edge can change by no more than $\Delta J$.}
\end{corollary}

\medskip

\begin{corollary}
\label{cor:cont}
{\rm The bond-averaged flexibility $\langle f\rangle$ of any ground state changes continuously when any coupling passes through its critical value.}
\end{corollary}

\section{Space-filling critical droplets}
\label{sec:sfcd}

For the remainder of this paper we will mostly be concerned with space-filling critical droplets which, as discussed in earlier papers~\cite{NS22,NS24c}, play a crucial role in determining which of several competing pictures of the low-temperature spin glass phase occurs in finite dimensions.  This role will be discussed further in Sect.~\ref{sec:discussion}.  In this section we prove a theorem (Theorem~\ref{thm:noexist}) which is one of the main results of this paper, namely that SFCD's cannot exist in the EA Ising model in any finite dimension. 

SFCD's have an important property: altering the coupling value of an edge in its boundary by a small amount (i.e., without causing a droplet flip) can change the flexibilities of a positive density of bonds in $\sigma$; when this occurs we will say that such a bond ``controls the flexibilities'' of the affected bonds. The next theorem shows that any bond in the boundary of an SFCD has a nonzero range of coupling values in which it controls the flexibilities of a positive density of bonds in $\sigma$. 

\medskip

\begin{theorem}
\label{thm:flexcontrol}
{\rm (Newman-Stein~\cite{NS22}, Theorem 6.3).  For $(J,\sigma)$ as in Theorem~\ref{thm:three}, and any bond~$b_0$ whose critical droplet~$D(b_0,\sigma)$ is space-filling in $\sigma$, there is an open nonempty interval of coupling values $J(b_0)$, with the critical value $J_c^\sigma(b_0)$ inside the interval, for which $J(b_0)$ controls the flexibilities of a positive density of bonds in ${\partial D}(b_0,\sigma)$.}
\end{theorem}

\medskip

This interval must be finite; there is an upper bound to how far it may extend.

\medskip

{\bf Definition}.  We say that a bond $b_{xy}$ (or its coupling $J(b_{xy})$) is {\it supersatisfied\/} in some fixed coupling realization $J$ if it is satisfied in every GSP.

It is not hard to see that a value of $\vert J(b_{xy})\vert$ above which $b_{xy}$ must be supersatisfied is 
\begin{equation}
\label{eq:ss}
\vert J(b_{xy})\vert\ge J^{\rm ub}(b_{xy})={\min}\Bigl(\sum_{z\ne y\atop |z-x|=1}\vert J_{xz}\vert,\sum_{u\ne x\atop |y-u|=1}\vert J_{uy}\vert\Bigr)\, .
\end{equation}
so the maximum length of the interval outside of which any bond must be supersatisfied is $2J^{\rm ub}$; however, some bonds could be supersatisfied outside a smaller interval.  An example of a situation where this occurs is shown in Fig.~\ref{fig:2D}, whose arrangement of couplings on a $2D$ square lattice has positive probability.  Similar examples can be constructed in higher dimensions.
\begin{figure}[h]
\centering
\includegraphics[scale=0.6]{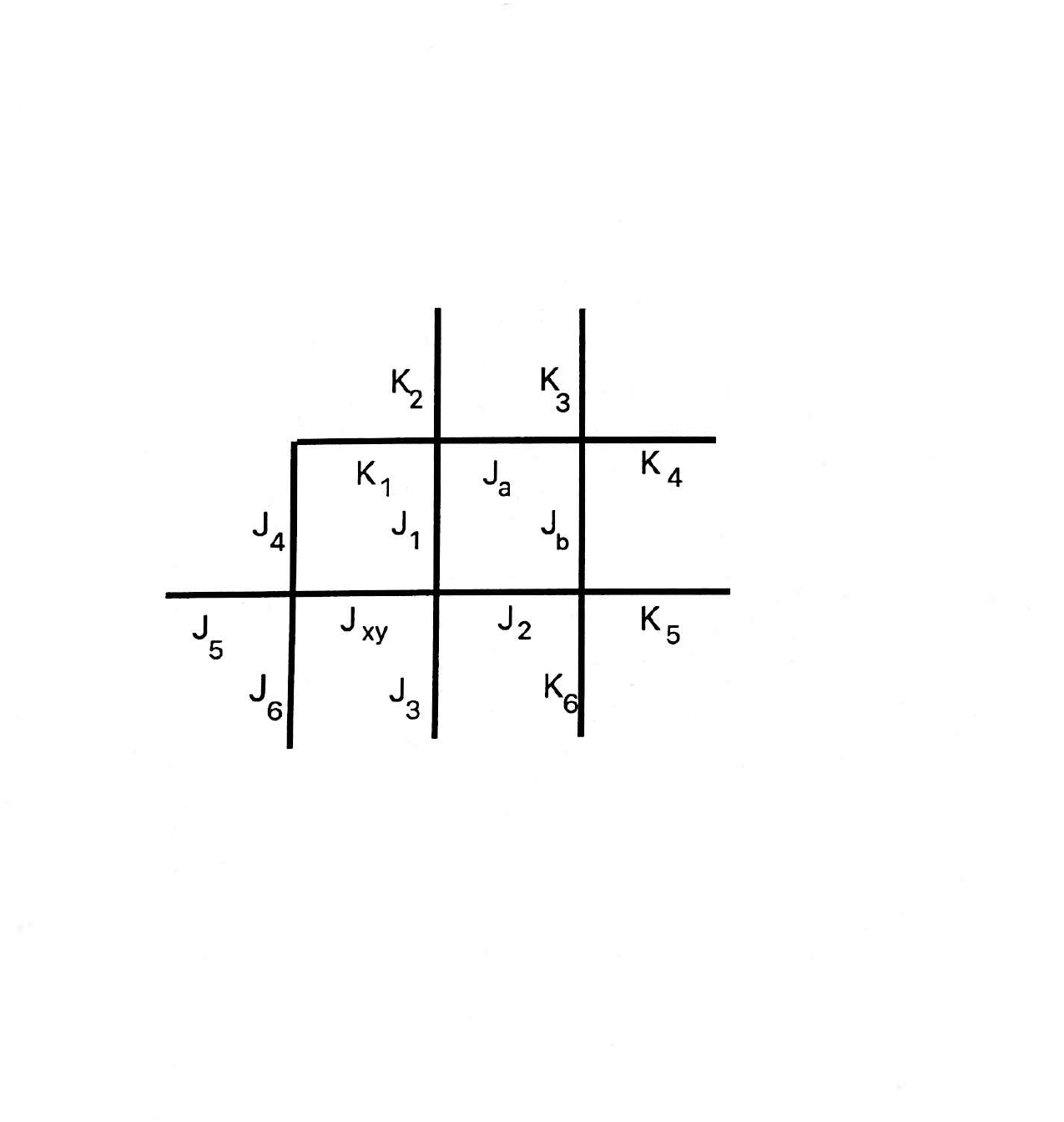}
\vskip -1.3in
\caption{Arrangement of couplings as described in the text.}
\label{fig:2D}
\end{figure}

In this particular coupling configuration the following relationships hold.  First, $\vert J_4\vert+\vert J_5\vert+\vert J_6\vert>\vert J_1\vert+\vert J_2\vert+\vert J_3\vert$, so for the bond $b_{xy}$ having coupling value $J_{xy}$,  $J^{\rm ub}=\vert J_1\vert+\vert J_2\vert+\vert J_3\vert$. We also take
\begin{equation}
\label{eq:couplings}
J_a\gg J_b\gg\{\vert J_1\vert,\vert J_2\vert,\vert K_1\vert,\vert K_2\vert,\vert K_3\vert,\vert K_4\vert,\vert K_5\vert,\vert K_6\vert\}
\end{equation}
so that $J_a$ and $J_b$ are both supersatisfied.  Finally, we take $\vert J_1\vert>\vert J_2\vert$ and ${\rm sgn} (J_1)=-{\rm sgn} (J_2)$.

It follows that in any GSP, $J_a$ and $J_b$ are both satisfied, while one of $J_1$, $J_2$ will be satisfied and the other unsatisfied.  From here it is not hard to see that, for the $J_{xy}$ in Fig.~\ref{fig:2D}, the length of the interval outside of which it is supersatisfied is at most $2\Big(\vert J_1\vert-\vert J_2\vert+\vert J_3\vert\Big)<2J^{\rm ub}$.

\medskip

A supersatisfied bond cannot be in the boundary of the critical droplet of any bond other than itself\footnote{With one exception: if for some $J$, $b_0$ is supersatisfied and the other bond in question (call it $b_1$)  is a neighbor that determines the range in which $b_0$ is supersatisfied (cf.~(\ref{eq:ss})), then $b_0$ could in principle belong to the critical droplet boundary of $b_1$. But this can only happen if $J(b_0)$ is no longer supersatisfied when $J(b_1)$ is sufficiently close to its critical value.}, and it cannot be in an interface between GSP's.  Although for fixed $J$ the range of the interval outside which a bond is supersatisfied depends on the bond,  we will omit the explicit bond-dependence when it is clear which bond is being referred to, and we will use $(J_{\rm lower},J_{\rm upper})$ to denote the interval outside of which a given bond is supersatisfied. In other words, for fixed $J$ a given coupling is {\it not\/} supersatisfied if it lies in the (bond-dependent) interval $(J_{\rm lower},J_{\rm upper})$; within this interval its corresponding bond can lie in the boundary of the critical droplet of a different bond, or in an interface between ground states.

\medskip

\begin{theorem}
\label{thm:gap}
{\rm (Newman-Stein~\cite{NS22}, Lemma~6.4).   If the critical droplet of a bond is spacefilling in a GSP~$\sigma$ then there is a nonzero gap between its critical value $J^\sigma_c$ and both $J_{\rm lower}$ and $J_{\rm upper}$; i.e., $J_{\rm lower}<J_c^\sigma<J_{\rm upper}$}.
\end{theorem}

\medskip

Using these results we now turn to the question of whether SFCD's can exist in any dimension.  Recall that Theorem~\ref{thm:three} asserts that any GSP in the support of a translation-covariant metastate will have either a positive density of bonds with SFCD's or none at all; therefore, in what follows when we say that a GSP has SFCD's, we will always mean a positive density of bonds has an SFCD.

We will need to consider several cases, given that the metastate~$\eta_J$ can be supported on a finite set of GSP's, a countable infinity of GSP's, or an uncountable infinity of GSP's. In the first two cases, all GSP's in the support of~$\eta_J$ have positive weight in the metastate, with the weights summing to one; in the third case, while there may be (a countable set of) GSP's with positive weight present (whose sum is then strictly less than one), there must always be an uncountable set of GSP's, each having zero weight but with the entire set having positive weight in~$\eta_J$.  In what follows we will consider in turn the cases of GSP's with positive weight in~$\eta_J$ and those with zero weight in~$\eta_J$.

\subsection{Positive weights}
\label{subsec:positive}

\begin{theorem}
\label{thm:finite}
{\rm  The metastate $\eta_J$ cannot be supported only on a finite set of GSP's with a) each having positive weight in~$\eta_J$ and b) at least one having edges whose critical droplets are space-filling.}
\end{theorem}

\medskip

{\bf Proof.} Let $N$ denote the number of GSP's in the support of $\eta_J$, with $1\le N<\infty$. Suppose first that $N=1$, and let $b_0$ denote a bond whose critical droplet is space-filling, and whose associated coupling~$J(b_0)$ in the single GSP~$\sigma$ has critical value $J^\sigma_c\in(J_{\rm lower},J_{\rm upper})$. Then by Theorem~\ref{thm:flexcontrol} there is an open interval of coupling values above (and below) $J^\sigma_c$ for which $J(b_0)$ controls the flexibilities of a positive density of bonds in ${\partial D}(b_0,\sigma)$ without causing a droplet flip.  Varying $J(b_0)$ toward $J_c^\sigma$ within this interval will therefore lower the average flexibility of $\sigma$.

There is an additional mechanism by which changing the flexibility of $J(b_0)$ can affect the average flexibility of $\sigma$. It might be the case that $b_0$ belongs to the critical droplets of a positive density of bonds not in ${\partial D}(b_0,\sigma)$. In~\cite{NS22} a bond with this property was said to exhibit $\sigma$-criticality of the second kind, but the definition there excluded bonds which already had SFCD's; we are broadening the definition here to include bonds which also have SFCD's.  

{\bf Definition.} If for some open interval of coupling values $J(b_{xy})$ a bond $b_{xy}$ belongs to the critical droplets of a positive density of bonds {\it not\/} in ${\partial D}(b_{xy},\sigma)$, we say it exhibits $\sigma$-{\it criticality of the second kind\/}.

If $\sigma$-criticality of the second kind were to occur, Lemmas~\ref{lem:remain} and~\ref{lem:flexlower} come into play, with the result that if $b_0$ has this property, it can only further lower the average flexibility of $\sigma$.

Varying $J(b_0)$ toward $J_c^\sigma$ (without crossing it)  will therefore lower the average flexibility of $\sigma$, and given that $\eta_J$ is supported on this single state, it will lower the average flexibility of $\eta_J$ as well, contradicting Theorem~\ref{thm:flexconst}.

We next turn to the case where $2\le N<\infty$. The proof for this case appears in Theorem~7.4 of~\cite{NS22}, but will be repeated here.  By assumption some bond $b_0$ has a SFCD in $1\le n\le N$ of the ground state pairs. We can relabel so that this subset of ground state pairs is $\sigma_1, \sigma_2, \ldots \sigma_n$ with $J_{c1}\ge J_{c2}\ge \ldots\ge J_{cn}$, where $J_{ci}$ is the critical value of $J(b_0)$ in ground state pair~$\sigma_i$. 

By Theorem~\ref{thm:gap} and the assumption that there is only a finite number of GSP's in~$\eta_J$, the intervals $\bigl[J_{c1},J_{\rm upper}\bigr]$ and $\bigl[J_{\rm lower},J_{cn}\bigr]$ have nonempty interiors. 
Choose $J^*$ so that $J_{\rm upper}>J^*>J_{c1}$.  It follows from Lemmas~\ref{lem:remain} and \ref{lem:flexlower} and Theorem~\ref{thm:gap} that lowering $J(b_0)$ from $J^*$ to $J_{c1}^+$ will lower the flexibilities in $\sigma_1$ of a positive density of bonds, and hence will change $P(f,\sigma_1)$. For all other ground states in the support of $\eta_J$, by Lemma~\ref{lem:flexlower} their average flexibilities will either be lowered or else remain unchanged.  Because $\sigma_1$ has positive weight in $\eta_J$, the average flexibility of $\eta_J$ will have changed which contradicts~Theorem~\ref{thm:flexconst}.~$\diamond$

\bigskip

{\bf Remark.} Theorems~\ref{thm:uv} and~\ref{thm:nowhere} below also rely on arguments in which $J(b_0)$ is varied toward the critical values of multiple GSP's without crossing any. Because $\sigma$-criticality of the second kind, should it occur, can only enhance the consequent lowering of the average flexibility of $\eta_J$, we will not explicitly note this in the proofs, but it should be understood.  In contrast, the proofs of Theorems~\ref{thm:dense} and~\ref{thm:atoms} do involve $J(b_0)$ crossing its critical value in some subset of GSP's, and the consequences of the possibility of $\sigma$-criticality of the second kind will be explicitly considered in those proofs.

\bigskip

\begin{theorem}
\label{thm:countable}
{\rm The metastate $\eta_J$ cannot be supported only on a countably infinite set of GSP's each of which have SFCD's.}
\end{theorem}

\medskip 

{\bf Proof.}  Let $\Sigma=\{-1,+1\}^{\mathbb{Z}^d}$ and let ${\cal M}_1(\Sigma)$ be the set of (regular Borel) probability measures on~$\Sigma$.  Consider a metastate $\eta_J$ of the form $\sum_\alpha W_\alpha\delta_{\Gamma^\alpha}$, where $\alpha$ is a positive integer labelling a GSP~$\Gamma^\alpha$ in the support of $\eta_J$, and $W_\alpha$ is the weight of $\Gamma_\alpha$ in~$\eta_J$; by assumption $\sum_\alpha W_\alpha=1$. 

If $T$ is a translation on $\mathbb{Z}^d$, then by translation-covariance of the metastate $\eta_{TJ}(\Gamma)=\eta_{J}(T^{-1}\Gamma)$, so the weight associated with $\Gamma^\alpha$ is the same as the weight of $T^{-1}\Gamma^\alpha$; i.e., the set of weights is translation-invariant as a function of $J$.  Therefore, the distribution of weights in the metastate is constant $\nu$-a.s., where $\nu$ is the distribution of the couplings.  

For every GSP whose weight $W_\alpha$ is distinct from all others, the index~$\alpha$ yields a measurable map of the couplings to ${\cal M}_1(\Sigma)$
\begin{equation}
\label{eq:mapsto}
J\mapsto \eta_J^\alpha :=\delta_{\Gamma_J^\alpha}\, .
\end{equation}
To see that $\eta_J^\alpha$ is a metastate, note that it is supported on GSP's, and has both translation- and coupling-covariance.  So $\eta_J^\alpha$ is a metastate supported on a single GSP. If multiple GSP's have the same weight, then by the same procedure one can construct a metastate containing only these GSP's, of which there must be a finite number.  In either case, the argument in the proof of Theorem~\ref{thm:finite} can be adapted here to show that by varying the coupling $J(b_0)$ the average flexibility of $\eta_J^\alpha$ also varies, leading to a contradiction.~$\diamond$

\medskip

The proof of Theorem \ref{thm:countable} leads to an immediate extension:

\medskip

\begin{theorem}
\label{thm:positive}
{\rm A GSP with positive weight in a zero-temperature metastate $\eta_J$ cannot have space-filling critical droplets.}
\end{theorem}

In the following section we consider scenarios where~$\eta_J$ is supported on a continuum of GSP's with zero weight.  In addition there may be ``mixed'' scenarios, where part of the support of $\eta_J$ is on GSP's with zero weight and part on GSP's with positive weight.  By Theorem~\ref{thm:positive}, any GSP with positive weight in $\eta_J$ cannot posses SFCD's, and therefore,  as $J(b_0)$ is varied, these cannot contribute to any change in the metastate flexibility distribution $P_J(f)$ defined in Theorem~\ref{thm:flexconst}.   We can therefore consider in what follows scenarios where $\eta_J$ is supported solely on GSP's with zero weight; the results obtained will apply equally to mixed scenarios.

\subsection{Zero weights}
\label{subsec:zero}

Suppose then that $\eta_J$ is supported entirely on an uncountable infinity of GSP's each having zero weight in $\eta_J$, and suppose at least part of the support of $\eta_J$ includes GSP's having a positive fraction of bonds whose critical droplets are space-filling (cf.~Theorem~\ref{thm:three}).  Because this set of GSP's is uncountable, whereas the set of edges in ${\mathbb E}^d$ is countable, there must exist a bond $b_0$ whose critical droplet is space-filling in a subset of GSP's with positive weight in $\eta_J$.  


\medskip 


We will be interested in the interval of values described in Theorem~\ref{thm:flexcontrol} for which $J(b_0)$ ``controls the flexibility'' (defined above Theorem~\ref{thm:flexcontrol}) of a positive density of bonds in $\partial D(b_0,\sigma)$, with all other couplings held fixed.   To study this, we define $a^+(b_0,\sigma)>0$ to be the largest value for which $J(b_0)\in\Big(J_c^\sigma,J_c^\sigma+a^+(b_0,\sigma)\Big)$ controls the flexibilities of a positive density of bonds in ${\partial D}(b_0,\sigma)$, and thus $J(b_0)\in\Big(J_c^\sigma,J_c^\sigma+a^+(b_0,\sigma)\Big)$ is a necessary and sufficient condition for a positive density of bonds in~$\partial D(b_0,\sigma)$ to share the same critical droplet~$D(b_0,\sigma)$.  Similarly, we define  $a^-(b_0,\sigma)>0$ to be the largest value for which $J(b_0)\in\Big(J_c^\sigma-a^-(b_0,\sigma),J_c^\sigma\Big)$ controls the flexibilities of a positive density of bonds in ${\partial D}(b_0,\sigma)$. Finally, we define $a(b_0,\sigma)=a^+(b_0,\sigma)+a^-(b_0,\sigma)$.  Equivalently, we can define these quantities as follows:

\medskip

{\bf Definition.} We will say that a coupling value $J(b_0)=J_0$ is {\it acceptable\/} if at $J_0$ the density of bonds~$\{b\in\partial D(b_0,\sigma):\ D(b,\sigma)=D(b_0,\sigma)\}$ is strictly greater than zero. Then define $a^+(b_0,\sigma)$ as $\sup\{{\hat a}:\ J_c^\sigma+{\hat a}\ {\rm is\ acceptable}\}$, $a^-(b_0,\sigma)$ as $\sup\{{\hat a}:\ J_c^\sigma-{\hat a}\ {\rm is\ acceptable}\}$, and $a(b_0,\sigma)=a^+(b_0,\sigma)+a^-(b_0,\sigma)$.

\medskip

When all couplings other than $J(b_0)$ are fixed, the condition $D(b,\sigma)=D(b_0,\sigma)$  restricts $J_0$ to lie in the interval~$(J_{\rm lower,}J_{\rm upper})$, as discussed just before the statement of Theorem~\ref{thm:gap}.

Theorem~\ref{thm:flexcontrol} then implies:

\medskip

\begin{theorem}
\label{thm:a}
{\rm For any GSP $\sigma$ and any $b_0$ whose critical droplet is space-filling in $\sigma$, $a^+(b_0,\sigma)>0$, $a^-(b_0,\sigma)>0$, and therefore $a(b_0,\sigma)>0$.}
\end{theorem}

The next lemma will be useful in what follows.

\medskip

\begin{lemma}
\label{lem:H}
{\rm $H(a):=\{\sigma:\ \sigma\ {\rm has}\ a(b_0,\sigma)\ge a\}$. Choose an interval $(c,d)\subseteq(J_{\rm lower},J_{\rm upper})$ and define $p$ as the metastate measure of $\sigma$'s with $J_c^\sigma\in(c,d)$. Then for any such $(c,d)$ with $p>0$ and any $k\in(0,1]$, there exists an $a>0$ such that at least a fraction $kp>0$ of $\sigma$'s with $J_c^\sigma\in(c,d)$ belongs to $H(a)$.}
\end{lemma}

\medskip

{\bf Proof.}  Note first that $H(0)$ contains the full set of $\sigma$'s with $J_c^\sigma\in (c,d)$ and therefore corresponds to $k = 1$; the fraction $k$ corresponding to the set of $\sigma$'s in $H(a)$ is monotonically nonincreasing as $a$ increases.  If the claim of the Lemma is false, then $H(a)$ for any $a>0$ corresponds to a fraction~$k=0$ of $\sigma$'s with $J_c^\sigma\in (c,d)$. This implies that a fraction $k=1$ of $\sigma$'s with $J_c^\sigma\in (c,d)$ have $a(b_0,\sigma)=0$, which contradicts~Theorem~\ref{thm:a}.~$\diamond$

{\bf Remark.}  Lemma~\ref{lem:H} also holds separately for $a^+(b_0,\sigma)$ (for some $a^+>0$ and $k^+>0$) and $a^-(b_0,\sigma)$ (for some $a^->0$ and $k^->0$).


\bigskip

With these definitions and results in hand, we now turn to the cases where $\kappa_J$ is supported on an uncountable infinity of GSP's.  Let ${\cal A}$ be the set of  $\sigma$'s with $D(b_0,\sigma)$ spacefilling; by the discussion above, $J_c^\sigma(b_0)\in(J_{\rm lower}(b_0),J_{\rm upper}(b_0))$ for all $\sigma\in{\cal A}$. From this point forward we will assume that each member of the set ${\cal A}$ has zero weight in $\eta_J$ but that the full set has positive weight in $\eta_J$.  We will consider two cases separately: Case I is where there is an open interval $(c,J_{\rm upper})$ for some $c<J_{\rm upper}$ and/or $(J_{\rm lower},d)$ for some $d>J_{\rm lower}$ in which either a) there are no $J_c^\sigma$'s or else (b) the set of $\sigma$'s with $J_c^\sigma\in(c,J_{\rm upper})$ and/or $J_c^\sigma\in(J_{\rm lower},d)$ has zero weight in $\eta_J$.  Case II is where the $J_c^\sigma$'s are dense over {\it both\/} intervals~$(J_{\rm lower},d)$ and $(c,J_{\rm upper})$, for some $c$ and $d$ with $J_{\rm lower}\le c<J_{\rm upper}$ and $J_{\rm lower}<d\le J_{\rm upper}$, and where both intervals individually have the property that the set of~$\sigma$'s with $J_c^\sigma$ in those intervals have positive weight in~$\eta_J$.

Before proceeding to a discussion of the two cases, we will need to consider the following possibility. Suppose that there exist $c$ and $d$ as just described for Case~II. It follows from the definition of $a^+(b_0,\sigma)$ that as $\sigma$ is varied so that $J_c^\sigma\to J_{\rm upper}$ from below, $a^+(b_0,\sigma)\to 0$, and similarly for $a^-(b_0,\sigma)$ as $J_{\rm lower}$ is approached from above (there is no contradiction with Theorem~\ref{thm:a}, given that by Theorem~\ref{thm:gap} there is zero probability for~$\sigma$ to have its $J_c^\sigma$ equal to either $J_{\rm upper}$ or $J_{\rm lower}$).  This raises the question of whether a similar event can occur at some other value(s) of $J(b_0)$; i.e., $a^+(b_0,\sigma)$ or $a^-(b_0,\sigma)$ goes to zero according to the following definition as $J_c^\sigma$ approaches some value $J^*\in(J_{\rm lower},J_{\rm upper})$ strictly between $J_{\rm min}$ and $J_{\rm max}$.

\medskip

{\bf Definition.} We will say that a value $J^*$ of the coupling $J(b_0)$ is a {\it null point\/} if $J_{\rm min}<J^*<J_{\rm max}$ and one (or both) of the following two possibilities occur(s): a) there exists~$J_1<J^*$ such that $a^+(b_0,\sigma)\le \vert J^*-J_c^\sigma\vert$ for almost all $\sigma$ having $J_c^\sigma\in(J_1,J^*)$, and/or b) there exists~$J_2>J^*$ such that $a^-(b_0,\sigma)\le \vert J^*-J_c^\sigma\vert$ for almost all $\sigma$ having $J_c^\sigma\in(J^*,J_2)$.

\medskip

\begin{lemma}
\label{lem:null}
{\rm Suppose that the set of $\sigma$'s with $J_c^\sigma\in(c,d)\subset(J_{\rm lower},J_{\rm upper})$ has positive $\eta_J$-measure.  Then null points for $a^+$ (resp.~$a^-$) cannot be dense in $(c,d)$.}
\end{lemma}

\medskip

{\bf Proof.}  Choose a GSP~$\sigma$ with $J_c^\sigma\in(c,d)$ and consider an open neighborhood $A_\epsilon$ of width $\epsilon>0$ containing $J_c^\sigma$. If null points are dense in $(c,d)$, then $A_\epsilon$ contains null points requiring $a^+(b_0,\sigma)\le\epsilon$. Because this is true for any $\epsilon>0$, it must be that $a(b_0,\sigma)=0$, violating Theorem~\ref{thm:a}. The same argument holds for $a^-$.~$\diamond$

\medskip

While the existence of null points seems unlikely, we have not ruled them out, and we will consider the possible effects of null points in Theorems~\ref{thm:uv},~\ref{thm:nowhere},~\ref{thm:dense},~\ref{thm:atoms}, and the Remark following~Theorem~\ref{thm:atoms}.

\subsubsection{Case I}
\label{subsec:nondense}

\begin{theorem}
\label{thm:uv}
{\rm Suppose that there is at least one open interval $(c,J_{\rm upper})$ with $c<J_{\rm upper}$ and/or $(J_{\rm lower},d)$ with $d>J_{\rm lower}$  in which either there are no $J_c^\sigma$'s or else the set of $\sigma$'s with $J_c^\sigma$ in one of the two intervals has zero weight in $\eta_J$; and furthermore suppose that the $J_c^\sigma$'s are dense in some adjoining interval $(u,c)$ with $J_{\rm lower}\le u<c$ and/or an adjoining interval $(d,v)$ with $d<v\le J_{\rm upper}$, and that in both cases their corresponding $\sigma$'s have positive weight in~$\eta_J$.  Moreover, suppose that $J(b_0)=c$ (if the relevant interval is $(c,J_{\rm upper})$) or $J(b_0)=d$ (if the relevant interval is $(J_{\rm lower},d)$) is not a null point. Then either none of the $\sigma\in{\cal A}$ has SFCD's, or at most a set of measure zero in~$\eta_J$ does.}
\end{theorem}

\medskip

{\bf Proof.}   Without loss of generality we can assume that it is the upper interval $(c,J_{\rm upper})$ which is devoid of $J_c^\sigma$'s, and the interval $(u,c)$ has $p>0$, where $p:=\eta_J\Big(\{\sigma:J_c^\sigma\in(u,c)\}\Big)$ like in Lemma~\ref{lem:H}, and furthermore suppose that the $J_c^\sigma$'s are dense within $(u,c)$.   By Lemma~\ref{lem:H} there exists an $a^+>0$ such that a positive fraction of $\sigma$'s with $J_c^\sigma\in(u,c)$ have $a^+(b_0,\sigma)\ge a^+$, and if there is no null point at $J(b_0)=c$, then $J(b_0)\in(c,c+a^+)$ will control the flexibilities of a positive density of bonds in a positive $\eta_J$-measure of GSP's with $J_c^\sigma\in(u,c)$ . With all other couplings held fixed, set $J(b_0)=c+\epsilon$, and choose $\epsilon\ll a^+$.  If we lower $J(b_0)$ from $c+\epsilon$ to $c+\epsilon/2$ no droplet flips occur in any GSP because $J(b_0)$ is within the gap in critical values, but the average flexibility is lowered in GSP's with $J_c^\sigma\in(c-a^+ +O(\epsilon),c)$, lowering in turn the average flexibility of $\eta_J$ and leading to a contradiction with Theorem~\ref{thm:flexconst}.

Next suppose that the intervals $(c,J_{\rm upper})$ and $(u,c)$ are defined as above except that now there may be $J_c^\sigma$'s in $(c,J_{\rm upper})$, but the set of $\sigma$'s with $J^\sigma_c\in (c,J_{\rm upper})$ has $\eta_J$-measure $p=0$. Then the same argument as that used above shows that such a scenario contradicts Theorem~\ref{thm:flexconst}, the only difference being that, while droplet flips in various $\sigma$'s occur when $J(b_0)$ is lowered from $c+\epsilon$ to $c+\epsilon/2$, these have no effect on the average metastate flexibility.~$\diamond$

\medskip

\begin{theorem}
\label{thm:nowhere}
{\rm Suppose as before that there is at least one open interval $(c,J_{\rm upper})$ with $c<J_{\rm upper}$ and/or $(J_{\rm lower},d)$ with $d>J_{\rm lower}$  in which either there are no $J_c^\sigma$'s or else the set of $\sigma$'s with $J_c^\sigma$ in one of the two intervals has zero weight in $\eta_J$. Suppose further that the set of GSP's with $J_c^\sigma$ in the open intervals~$(u,c)$ and $(d,v)$, with $u$ and $v$ defined as in Theorem~\ref{thm:uv}, both have positive $\eta_J$-weight but now have the property that there is no open subset of either~$(u,c)$ or $(d,v)$ in which the $J_c^\sigma$'s are dense.
Moreover, suppose as in Theorem~\ref{thm:uv} that $J(b_0)=c$ or $J(b_0)=d$ is not a null point. Then either none of the $\sigma\in{\cal A}$ has SFCD's, or at most a set of measure zero in~$\eta_J$ does.}
\end{theorem}

\medskip
 
{\bf Proof.} As before, we focus on the upper interval $(c,J_{\rm upper})$ which is devoid of $J_c^\sigma$'s, and the adjoining interval $(u,c)$ which has positive $\eta_J$-weight.  By Lemma~\ref{lem:H} there exists an $a^+>0$ such that a positive fraction of $\sigma$'s with $J_c^\sigma\in(u,c)$ have $a^+(b_0,\sigma)\ge a^+$, and if there is no null point at $J(b_0)=c$, then as in the proof of Theorem~\ref{thm:uv}, $J(b_0)\in(c,c+a^+)$ will control the flexibilities of a positive density of bonds in a positive $\eta_J$-measure of GSP's with $J_c^\sigma\in(u,c)$.   Now let $J(b_0)$ reside in a gap (empty of $J^\sigma_c$'s) just above $c$ and lower it by an amount $\epsilon$ sufficiently small such that $J(b_0)$ does not cross $c$ (so it crosses no $J_c^\sigma$'s). During this process no droplet flips occur in any GSP because $J(b_0)$ is within the gap in critical values, but the average flexibility is lowered in GSP's with  $J_c^\sigma\in(c-a^+ +O(\epsilon),c)$, which contradicts Theorem~\ref{thm:flexconst}.  As in the proof of Theorem~\ref{thm:uv}, the argument is essentially the same if $(c,J_{\rm upper})$ contains $J_c^\sigma$'s but the set of GSP's with $J_c^\sigma\in(c,J_{\rm upper})$ has zero $\eta_J$-weight. ~$\diamond$

\medskip

{\bf Remark.} The proof of Theorem~\ref{thm:nowhere} implicitly assumes that any open subset of $(u,c)$ has positive $\eta_J$-weight. Of course it is also possible that there exists some $w$ with $u<w<c$ such that the set of GSP's with $J_c^\sigma\in(w,c)$ has zero $\eta_J$-weight (and now the set of GSP's with $J_c^\sigma\in(u,w)$ has positive $\eta_J$-weight). In this case one simply repeats the above argument using the subset $(u,w)$ in place of $(u,c)$.

{\bf Remark.} The results of Theorems~\ref{thm:uv} and~\ref{thm:nowhere} can be extended to the case where $c$ or $d$ is a null point; we will return to this case following the proof of Theorem~\ref{thm:atoms} below.

\bigskip


\subsubsection{Case II}
\label{subsec:everywhere}

The remaining case is where the set ${\cal A}$ has the following properties: a) the set of $\sigma\in{\cal A}$, and their corresponding $J_c^\sigma$'s in $(J_{\rm lower},J_{\rm upper})$, is uncountable; b) each $\sigma\in{\cal A}$ has zero weight in~$\eta_J$; c) the full set of $\sigma\in{\cal A}$ has positive measure in~$\eta_J$; d) the $J_c^\sigma$'s are dense over {\it both\/} intervals~$(c,J_{\rm upper})$ and $(J_{\rm lower},d)$, for some $c$ and $d$ with $J_{\rm lower}\le c<J_{\rm upper}$ and $J_{\rm lower}<d\le J_{\rm upper}$, and e) both intervals individually have the property that the set of~$\sigma$'s with $J_c^\sigma$ in those intervals have positive weight in~$\eta_J$.  

Before proceeding it is helpful to introduce a probability measure $\rho(J_0)$ (where $J_0=J(b_0)$) whose domain is $J_0\in(J_{\rm lower},J_{\rm upper})$ and with the following properties: 1) $\rho(J_0)\ge 0$; 2) $\int_{J_{\rm lower}}^{J_{\rm upper}}\rho(J_0)\ dJ_0=1$;  and 3)  $\int_{a}^{b}\rho(J_0)\ dJ_0>0$ for all $J_{\rm lower}<a<b<J_{\rm upper}$.  Here $\int_{a}^{b}\rho(J_0)\ dJ_0$ is the fraction of ground states with $J_c$ in the interval $(a,b)$ relative to ground states with $J_c$ anywhere within the entire interval $(J_{\rm lower},J_{\rm upper})$. 

Even though we are now assuming an atomless continuum of ground states in~$\eta_J$, {\it a priori\/} it might be that $\rho(J_0)$ has atoms with positive weight in~$\eta_J$; this would occur if a set of $\sigma$'s with positive weight in $\eta_J$ have the same value of $J_c^\sigma(b_0)$.   As will be seen in the proof of Theorem~\ref{thm:dense} below, the only case that will need to be considered is one in which these atoms are dense throughout~$(c,J_{\rm upper})$ and $(J_{\rm lower},d)$.  We return to this after stating and proving Theorem~\ref{thm:dense}, which considers the case in which the critical values of $J_0$ form an atomless continuum and are dense in the intervals $(c,J_{\rm upper})$ and $(J_{\rm lower},d)$.

\medskip

\begin{theorem}
\label{thm:dense}
{\rm Suppose the set ${\cal A}$ along with $\eta_J$ and $\rho$ have the following properties: a) the set of $\sigma\in{\cal A}$, and their corresponding $J_c^\sigma$'s in $(J_{\rm lower},J_{\rm upper})$, is uncountable; b) each $\sigma\in{\cal A}$ has zero weight in~$\eta_J$; c) the full set has positive measure in~$\eta_J$; d) there are no atoms in $\rho(J_0)$; e) the set of GSP's with $J_c^\sigma\in(c,J_{\rm upper})$ has positive measure in $\eta_J$, and similarly for the set of GSP's with $J_c^\sigma\in(J_{\rm lower},d)$; and f) the $J_c^\sigma$'s are dense in both $(c,J_{\rm upper})$ and $(J_{\rm lower},d)$ .  If these conditions are satisfied, there are no SFCD's for a.e.~$\sigma$ in~${\cal A}$.}
\end{theorem}

\medskip

{\bf Proof.}   By Lemma~\ref{lem:null} there is a $c<J_{\rm upper}$ such that the interval $(c,J_{\rm upper})$ has no null points; we confine ourselves to this interval.  Consider the behavior of the metastate average of the flexibility $\langle f\rangle_{\eta_J}$.  We wish to study the change in the mean flexibility $\Delta\langle f\rangle(\epsilon)$ when $J_0$ changes from $J_{\rm upper}^+$ to $J_{\rm upper}-\epsilon$.   The net flexibility change of any GSP with $J_c\in(J_{\rm upper}-\epsilon,J_{\rm upper})$ may be either positive or negative during this process: as shown in Lemma~\ref{lem:flexlower}, in a GSP~$\sigma$ as $J(b_0)$ moves toward $J^\sigma_c$ from above, the flexibility of any coupling can only decrease or remain unchanged; after $J(b_0)$ crosses $J^\sigma_c$ and continues to decrease, the flexibility in $\sigma$ of any coupling can only increase or remain the same. 

An upper bound on the positive change of the mean flexibility can be obtained using Lemma~\ref{lem:flexlower} and Corollary~\ref{cor:flexchange} by assuming that the flexibility of {\it every\/} bond (i.e., all of $\mathbb{E}^d$) in GSP's in which $J(b_0)$ has passed through their respective $J_c$'s has the maximum possible increase~$\epsilon$; these correspond to ground states with $J_c\in(J_{\rm upper}-\epsilon,J_{\rm upper})$. We then have
\begin{equation}
\label{eq:2}
\Delta^{+}\langle f\rangle\le \epsilon\int_{J_{\rm upper}-\epsilon}^{J_{\rm upper}}\rho(J_0)\ dJ_0\, .
\end{equation}
The superscript $+$ on the LHS denotes that the change is only for ground states in the interval used above. 

We next examine the change in flexibility of GSP's with $J_c<J_{\rm upper}-\epsilon$.  Again, using Lemma~\ref{lem:flexlower} and Corollary~\ref{cor:flexchange}, for all of these the average flexibility can only decrease or remain unchanged as $J_0$ is lowered to $J_{\rm upper}-\epsilon$. Using Lemma~\ref{lem:H}, there exists $a^+>0$ and corresponding $k^+>0$ such that a fraction~$\ge k^+$ of GSP's with $J_c\in[J_{\rm upper}-\epsilon-a^+, J_{\rm upper}-\epsilon]$ will have their average flexibilities lowered (here we've chosen $\epsilon\ll a^+$). GSP's with $J_c^\sigma$ outside this range may have their average flexibilities lowered as well, so by neglecting these we will obtain a lower bound for the magnitude of the overall decrease in flexibility.

In a given $\sigma$ with $J^\sigma_c<J_{\rm upper}-\epsilon$, any edge in $\partial D(b_0,\sigma)$ whose flexibility is controlled by $J_0$ throughout the interval $J_0\in(J_{\rm upper}-\epsilon,J_{\rm upper})$ will have its flexibility decreased by $\epsilon$. Similarly, if $b_0$ exhibits $\sigma$-criticality of the second kind in a positive fraction of $\sigma$'s with $J_c^\sigma\in (J_{\rm upper}-a^+-\epsilon, J_{\rm upper}-\epsilon)$, then any bonds not in ${\partial D}(b_0,\sigma)$ but whose critical droplet includes $b_0$ will similarly have their flexibilities decreased by $\epsilon$. The total flexibility decrease in each GSP depends on how many bonds share the critical droplet $D(b_0,\sigma)$ and how many bonds not in ${\partial D}(b_0,\sigma)$ have a  critical droplet whose boundary includes $b_0$.  By Lemma~\ref{lem:H}, for any interval $(c,d)\subseteq(J_{\rm lower},J_{\rm upper})$, there must exist~$q>0$ and $p>0$ such that in a fraction $p$ of the ground states with $J_c\in (c,d)$ the density of bonds in $\partial D(b_0,\sigma)$ whose flexibility is controlled by $J_0$ is greater than $q$. By ignoring the additional contribution to the lowering of the average flexibility of $\eta_J$ due to $\sigma$-criticality of the second kind, we have the following bound for the change in average flexibility of $\eta_J$ due to ground states with $J^\sigma_c<J_{\rm upper}-\epsilon$:
\begin{equation}
\label{eq:2a}
\Delta^{-}\langle f\rangle\le -qp\epsilon\int_{J_{\rm upper}-a^+-\epsilon}^{J_{\rm upper}-\epsilon}\rho(J_0)\ dJ_0\, .
\end{equation}
We therefore have for the overall change in average metastate flexibility:
\begin{equation}
\label{eq:3}
\Delta\langle f\rangle=\Delta^{+}\langle f\rangle+\Delta^{-}\langle f\rangle\le \epsilon\Biggl(\int_{J_{\rm upper}-\epsilon}^{J_{\rm upper}}\rho(J_0)\ dJ_0-qp\int_{J_{\rm upper}-a^+-\epsilon}^{J_{\rm upper}-\epsilon}\rho(J_0)\Biggr)\, .
\end{equation}

The first term inside the parentheses on the RHS can only decrease as $\epsilon$ decreases, and in fact goes to zero as $\epsilon\to 0$.  The magnitude of the second term inside the parentheses on the other hand, is bounded away from zero.  Therefore, for some sufficiently small $\epsilon$, the total change in flexibility is negative.


Consequently the average flexibility of $\eta_J$ will be lowered by this process, leading to a contradiction. $\diamond$

\bigskip

We now return to the case where $\rho(J_0)$ has atoms.  The proof of Theorem~\ref{thm:dense} fails only if these atoms form a dense set in both $(J_{\rm upper} - c_1,J_{\rm upper})$ and $(J_{\rm lower},J_{\rm lower}+c_2)$ for some $c_1,c_2>0$.  Suppose this is the case. By Lemma~\ref{lem:null} there is $c_3<J_{\rm upper}$ such that the interval $(c_3,J_{\rm upper})$ has no null points; we confine ourselves to this interval.  By Lemma~\ref{lem:H}, there exists $a^+>0$ such that a positive fraction of $\sigma$'s in $(J_{\rm upper}-a^+,J_{\rm upper})$ have the property that $J(b_0)$ controls the flexibilities of a positive fraction of bonds in~${\partial D}(b_0,\sigma)$.

For the moment we consider only changes in flexibilities of bonds that lie in ${\partial D}(b_0,\sigma)$.  Divide the interval $(J_{\rm upper}-a^+,J_{\rm upper})$ into two subintervals $(J_{\rm upper}-a^+,J_{\rm upper}-ka^+)$ and $(J_{\rm upper}-ka^+,J_{\rm upper})$ with $0<k<1$, and with $k$ chosen as follows. As $J_0=J(b_0)$ is lowered to some value $J_1$ below $J_{\rm upper}$, the average flexibilities in $\sigma$'s with $J_c^\sigma>J_1$ will either increase or decrease, while those with $J_c^\sigma<J_1$ can only decrease.  The maximum increase of flexibility $\Delta^+(k)$ when $J_0$ is lowered from $J_{\rm upper}$ to $J_{\rm upper}-ka^+$ can then be bounded by
\begin{equation}
\label{eq:+}
\Delta^+(k)\le ka^+\int_{J_{\rm upper}-ka^+}^{J_{\rm upper}} \rho(J_0)\ dJ_0\, .
\end{equation}
If $\rho(J_0)$ has atoms then $\Delta^+(k)$ will make discontinuous jumps at various values of $J_0$.

Before discussing the decrease $\Delta^-(k)$ of the flexibilities of bonds with $J_c^\sigma\in(J_{\rm upper}-a^+,J_{\rm upper}-ka^+)$ when $J_0$ is lowered from $J_{\rm upper}$ to $J_{\rm upper}-ka^+$, we introduce a few new quantities.  Let $p(b_0,\sigma)$ be the density of bonds in ${\partial D}(b_0,\sigma)$ and $r(b_0,\sigma)$  be the fraction of bonds in ${\partial D}(b_0,\sigma)$ whose flexibilities are controlled by $J_0$ when $J_0=J_{\rm upper}$.  Because ${\partial D}(b_0,\sigma)$ is space-filling by assumption, $p(b_0,\sigma)>0$. By Lemma~\ref{lem:H}, a positive $\eta_J$-measure of $\sigma$'s with $J_c^\sigma\in(J_{\rm upper}-a^+,J_{\rm upper}-ka^+)$ will have $r(b_0,\sigma)>0$, and moreover the fraction of bonds in ${\partial D}(b_0,\sigma)$ whose flexibilities are controlled by $J_0$ can only {\it increase\/} as $J_0$ is lowered from $J_{\rm upper}$ to $J_{\rm upper}-ka^+$. We note that the flexibility of any bond whose own critical droplet switches to ${\partial D}(b_0,\sigma)$ as $J_0$ is lowered (these are the Type~3 bonds introduced in the proof of Lemma~\ref{lem:flexlower}) will have a decrease in flexibility greater than zero but strictly smaller than $ka^+$.  Ignoring the contribution of such bonds, and also (as in the proof of Theorem~\ref{thm:dense})  ignoring additional contributions to flexibility decrease due to $\sigma$-criticality of the second kind, leads to a lower bound for the magnitude of the decrease of flexibility arising from GSP's with $J_c^\sigma\in(J_{\rm upper}-a^+,J_{\rm upper}-ka^+)$ as $J_0$ is lowered from $J_{\rm upper}$ to $J_{\rm upper}-ka^+$:
\begin{equation}
\label{eq:-}
\vert\Delta^-(k)\vert\ge ka^+\int_{J_{\rm upper}-a^+}^{J_{\rm upper}-ka^+} dJ_0\int d\kappa_J(\sigma) {\bf 1}_{J_c^\sigma=J_0}\ p(b_0,\sigma)r(b_0,\sigma)\ ,
\end{equation}
where $\Delta^-(k)<0$, ${\bf 1}$ is the indicator function and the subscript~$J$ in $d\kappa_J(\sigma)$ indicates all couplings are fixed except for $J(b_0)=J_0$. (Varying $J(b_0)$ has no effect on $p(b_0,\sigma)$, $r(b_0,\sigma)$ or any of the $J_c^\sigma$'s.)

When $k=0$, $\int_{J_{\rm upper}-ka^+}^{J_{\rm upper}} \rho(J_0)\ dJ_0=0$ and $\int_{J_{\rm upper}-a^+}^{J_{\rm upper}-ka^+} dJ_0\int d\kappa_J(\sigma) {\bf 1}_{J_c^\sigma=J_0} p(b_0,\sigma)r(b_0,\sigma)$ is at its maximum. As $k$ increases $\Delta^+(k)$ increases from 0 and $\vert\Delta^-(k)\vert$ decreases from its maximum.  Because of the atoms in $\rho(J)$ there will be jumps in both as $k$ is varied. There must then be some $k_0>0$ above which $\Delta^+(k)>\vert\Delta^-(k)\vert$ and below which $\Delta^+(k)<\vert\Delta^-(k)\vert$. We then choose $k=k_0-\epsilon$ with $0<\epsilon<k_0$. But this then violates Theorem~\ref{thm:flexconst}, which requires $\Delta^+(k)=\vert\Delta^-(k)\vert$ for any value of $k$. We have therefore shown

\medskip

\begin{theorem}
\label{thm:atoms}
{\rm Given the conditions stated in Theorem~\ref{thm:dense}, with the exception that now $\rho(J_0)$ has atoms, there are no SFCD's for a.e.~$\sigma$ in~${\cal A}$.}
\end{theorem}

\medskip

{\bf Remark.} We now return to Theorems~\ref{thm:uv} and~\ref{thm:nowhere}, and suppose for each that there is a null point at $J(b_0)=c$ (or $d$), where $c$ and $d$ are as in those theorems.  The same arguments as those used in the proofs of Theorem~\ref{thm:dense} (if $\rho(J_0)$ has no atoms) or Theorem~\ref{thm:atoms} (if $\rho(J_0)$ has atoms) can be used to rule out the existence of SFCD's in those situations, where $c$ plays the same role as $J_{\rm upper}$ and/or $d$ plays the same role as $J_{\rm lower}$.



\bigskip

Combining Theorems~\ref{thm:finite}-\ref{thm:positive}, \ref{thm:uv}-\ref{thm:nowhere}, and \ref{thm:dense}-\ref{thm:atoms} we have:

\medskip

\begin{theorem}
\label{thm:noexist}
{\rm Space-filling critical droplets do not exist for a.e.~ground state chosen from~$\eta_J$.}
\end{theorem}

\medskip

{\bf Remark.}  Theorem~\ref{thm:noexist} does not rule out the possible presence of zero-density critical droplets that overturn an infinite set of spins.  

\medskip

In the next section we examine an important consequence of Theorem~\ref{thm:noexist}.

\section{Fluctuations in ground state energy differences}
\label{sec:energyflucs}

In this and the following sections we confine ourselves to a zero-temperature periodic boundary condition (PBC) metastate $\kappa_J$, defined in the paragraph following Eq.~(\ref{eq:cc}). Because $\kappa_J$ is a member of the more general class~${\cal N}_J$, Theorem~\ref{thm:noexist} applies, so the ground states in the support of $\kappa_J$ have no SFCD's.

It was proved in~\cite{NS01c} that if a zero-temperature~$\kappa_J$ is supported on multiple GSP's (recall that all ground states in the support of $\kappa_J$ come in globally spin-reversed pairs), then these GSP's must be mutually incongruent~\cite{HF87,FH87}, i.e.~their relative interface has positive density in the edge set ${\mathbb E}^d$.  This result was extended to pure states at positive temperature in Theorem~5.2 of~\cite{NS24b}.  More precisely, define the edge overlap between two distinct GSP's (or pure state pairs at positive temperature) $\alpha$ and $\beta$ as follows. If $\mathbb{E}_L=\mathbb{E}_{\Lambda_L}$ denotes the edge set within the volume $\Lambda_L$, then the edge overlap between $\alpha$ and $\beta$ is defined as
\begin{equation}
\label{eq:overlap}
q^{(e)}_{\alpha\beta}=\lim_{L\to\infty}\frac{1}{d\vert\Lambda_L\vert}\sum_{\langle xy\rangle\in \mathbb{E}_L}\langle\sigma_x\sigma_y\rangle_\alpha\langle\sigma_x\sigma_y\rangle_\beta\, .
\end{equation}



The limit in~(\ref{eq:overlap}) exists by the spatial ergodic theorem~\cite{Wiener39,Readcomm}.  (Even if this were not the case, the limit can be replaced by the lim sup, which is guaranteed to exist~\cite{Royden10}, and this would then serve as the definition of the edge overlap.   In what follows we will use existence of the limit of the edge overlap, but note that even if this were not the case, the arguments would still go through using the lim sup.)  Two ground (or pure) state pairs $\alpha$ and $\beta$ are incongruent if $q_{\alpha\beta}^{(e)}<1$ (at zero temperature) or $q_{\alpha\beta}^{(e)}<q_{\alpha\alpha}^{(e)}$ (at positive temperature).  ($q_{\alpha\alpha}^{(e)}$ is the equivalent of the EA order parameter $q_{EA}$ for bond variables, and has the same value for all pure states in the positive-temperature metastate~\cite{NRS23a,Read24}.)

In~\cite{NS24a,NS24b} it was proved (cf.~Lemma~4.1 in~\cite{NS24b}) that at positive temperature the edge overlap $q^{(e)}_{\alpha\beta}$ is invariant under a change of finitely many couplings, and this served as an essential ingredient in the main result of those papers, namely that in the positive-temperature PBC metastate, the free energy difference between any two incongruent pure states in its support has variance which scales with the volume.   This result was confined to positive temperature because it relied on the invariance of the edge overlap with respect to finite changes in the coupling realization. The possible existence of SFCD's prevented the extension to zero temperature, but now that such critical droplets have been ruled out, the result can be extended to zero temperature.  

To see this, suppose for the sake of argument that SFCD's do exist. Consider a GSP $\alpha$ taken from the support of $\kappa_J$ and suppose that it has SFCD's, and as before let $b_0$ be a bond whose critical droplet boundary ${\partial D}(b_0,\alpha)$ in $\alpha$ has positive density.  Similarly, let $\alpha^>$ denote the GSP when $J(b_0,\alpha)>J_c^\alpha$ and $\alpha^<$ denote the GSP when $J(b_0,\alpha)<J_c^\alpha$.  If the edge overlap $q^{(e)}_{\alpha^>\alpha^<}$ exists, then the density of ${\partial D}(b_0,\alpha)$ is well-defined: it is simply $\frac{1}{2}(1-q^{(e)}_{\alpha^>\alpha^<})$.  However, even though edge overlaps between incongruent GSP's taken from $\kappa_J$ exist, the same is not necessarily true for $q^{(e)}_{\alpha^>\alpha^<}$, since $\alpha^>$ and $\alpha^<$ are GSP's for coupling realizations that differ by a single coupling.

If so, one can instead do the following: in addition to $\alpha$, choose a second GSP $\beta$ from $\kappa_J$ with $J_c^\alpha(b_0)\ne J_c^\beta(b_0)$. Such a~$\beta$ must exist, because if not, the distribution of $J_c^\sigma$'s in $\rho(J_0)$ is a single $\delta$-function, which is ruled out by arguments similar to those in the proof of Theorem~\ref{thm:finite}.  Proceeding, one then lowers the coupling value $J(b_0)$ from $J_c^\alpha(b_0)+\epsilon$ to $J_c^\alpha(b_0)-\epsilon$; because $J_c^\alpha(b_0)\ne J_c^\beta(b_0)$ one can always find an $\epsilon>0$ such that $J_c^\beta(b_0)$ is not crossed, so $\beta$ is unaffected by the change in coupling value. One then compares $q^{(e)}_{\alpha^>\beta}$ (evaluated at  $J_c^\alpha(b_0)+\epsilon$) to $q^{(e)}_{\alpha^<\beta}$ (evaluated at  $J_c^\alpha(b_0)-\epsilon$). 

To extend the results of~\cite{NS24a,NS24b} to zero temperature, we require $q^{(e)}_{\alpha^>\beta}=q^{(e)}_{\alpha^<\beta}$ for all $\alpha$ and $\beta$ chosen from $\kappa_J$. Suppose that this is not the case, i.e.,  $q^{(e)}_{\alpha^>\beta}\ne q^{(e)}_{\alpha^<\beta}$.  Then
\begin{eqnarray}
\label{eq:density}
0<\Bigl\vert q^{(e)}_{\alpha^>\beta} - q^{(e)}_{\alpha^<\beta}\Bigr\vert=\lim_{L\to\infty}\frac{1}{d\vert\Lambda_L\vert}\Biggl\vert\sum_{\langle xy\rangle\in \mathbb{E}_L}\Bigl(\langle\sigma_x\sigma_y\rangle_{\alpha^>}-\langle\sigma_x\sigma_y\rangle_{\alpha^<}\Bigr)\langle\sigma_x\sigma_y\rangle_\beta\Biggr\vert\nonumber\\
\le \lim_{L\to\infty}\frac{1}{d\vert\Lambda_L\vert}\sum_{\langle xy\rangle\in \mathbb{E}_L}\Biggl\vert\langle\sigma_x\sigma_y\rangle_{\alpha^>}-\langle\sigma_x\sigma_y\rangle_{\alpha^<}\Biggr\vert\Biggl\vert\langle\sigma_x\sigma_y\rangle_\beta\Biggr\vert\nonumber\\
=2\mu\Bigl({\partial D}(b_0,\alpha)\Bigr)\, ,
\end{eqnarray}
where $\mu\Bigl({\partial D}(b_0,\alpha)\Bigr)$ is the density of~${\partial D}(b_0,\alpha)$ if it exists.

This shows that $\frac{1}{2}\Bigl\vert q^{(e)}_{\alpha^>\beta} - q^{(e)}_{\alpha^<\beta}\Bigr\vert$ provides a lower bound on the upper density of ${\partial D}(b_0,\alpha)$.  (To get the best lower bound in either case, one looks for the GSP $\beta$ in the support of $\kappa_J$ that maximizes $\vert q^{(e)}_{\alpha^>\beta} - q^{(e)}_{\alpha^<\beta}\vert$.)  Moreover, the equality in the first line of~(\ref{eq:density}) implies that if the upper density of ${\partial D}(b_0,\alpha)$ is zero, then $q^{(e)}_{\alpha^>\beta} = q^{(e)}_{\alpha^<\beta}$.

The preceding discussion proves the following lemma:

\medskip

\begin{lemma}
\label{lem:qab}
{\rm Any change in $q^{(e)}_{\alpha\beta}$ is directly related to the (positive) density of critical droplet boundaries in $\alpha$ and $\beta$; if neither contain SFCD's  then $q^{(e)}_{\alpha\beta}$ is unchanged when finitely many couplings are varied.}
\end{lemma} 

\medskip

The result in~\cite{NS24a,NS24b} on free energy difference fluctuations between pure states followed from the construction of a new type of object, the so-called {\it restricted metastate\/} ${\kappa}^{p,\delta}_{J,\omega}$, which at zero temperature can be defined as follows: first, choose a GSP~$\omega$ from the distribution~$\kappa_J(\omega)$, where as before $\kappa_J$ is a zero-temperature PBC metastate.  (In order to construct a new metastate, every GSP $\omega$ in~$\kappa_J$ needs to be considered as a possible reference ground state.) We also choose an interval $(p-\delta,p+\delta)$ with $p\in[-1,1]$, $\delta\ge 0$ and 
\begin{equation}
\label{eq:delta}
\delta\ll\begin{cases}{\rm min}(p,1-p) & p>0\, ,\\
{\rm min}(1+p,-p)&p<0\, ,\\
1& p=0\, . 
\end{cases}
\end{equation}

In~\cite{NS24b} two constructions were presented which may give rise to different restricted metastates, but each of which satisfies the properties required of a metastate.  We refer the reader to~\cite{NS24b} for details, but give a brief summary here. Before proceeding we note that because periodic boundary conditions are used to construct~$\kappa_J$, every Gibbs state~$\Gamma$ in the metastate will be a mixture of two or more pure states at positive temperature, while at zero temperature each will be an equal mixture of a ground state and its spin-reversed counterpart.  In Construction~1 one chooses a Gibbs state~$\Gamma$ from $\kappa_J$ and a reference pure (or ground) state $\omega$ from~$\Gamma$.  Next one considers every pure or ground state~$\alpha$ in~$\kappa_J$ and retains only those whose edge overlap $q^{(e)}_{\alpha\omega}$ with $\omega$ is within the predetermined restricted interval $[p-\delta,p+\delta]$.  The remaining pure or ground states are discarded and the weights of the remaining states renormalized so that their probabilities add to one.  In Construction~2, for a given $\omega$ one retains only the $\Gamma$ from which $\omega$ was chosen, and then follows the procedure of Construction~1 for each of the pure or ground states in $\Gamma$ alone.

If $p$ is chosen so that the resulting set of GSP's is nonempty, the resulting object is a~$(p,\delta)$-restricted measure ${\kappa}^{p,\delta}_{J,\omega}$ on ground state pairs.  (We note that in Construction~2 the parameter~$p$ is chosen from the support of the edge overlap measure within $\Gamma$ alone.)  Then ${\kappa}^{p,\delta}_{J,\omega}$ as constructed above satisfies the three conditions for a translation-covariant metastate, but now depending on both $J$ and $\omega$, by reasoning similar to that in the proof of Theorem~7.1 of~\cite{NS24b}.  

It was shown in~\cite{NS24b} (Theorem~6.1) that two restricted metastates ${\tilde\kappa}^{p_1,\delta}_{J,\omega}$ and ${\tilde\kappa}^{p_2,\delta}_{J,\omega}$ with $0\le p_1<p_2\le 1$ and $0\le\delta< {\rm min}(p_1,p_2-p_1, 1-p_2)$ are themselves incongruent, in the following sense: for any edge~$(x,y)$
\begin{equation}
\label{eq:incong2}
(\nu\times\kappa_J)\Bigl\{(J,\omega):{\kappa}^{p_1,\delta}_{J,\omega}\big(\langle\sigma_x\sigma_y\rangle_\alpha\big)\neq {\kappa}^{p_2,\delta}_{J,\omega}\big(\langle\sigma_x\sigma_y\rangle_\alpha\big)\Bigr\}>0\, ,
\end{equation}
where $\nu\times\kappa_J$ denotes $\nu(dJ)\kappa_J(d\omega)$.

Now consider the energy difference between two incongruent infinite-volume  GSP's $\alpha$ and $\beta$ chosen from $\kappa_J$:
\begin{equation}
\label{eq:ed}
{\cal E}_L(J,\alpha,\beta)={\cal H}_{\Lambda_L, J}(\alpha)-{\cal H}_{\Lambda_L, J}(\beta)
\end{equation}
where, using~(\ref{eq:EA}), ${\cal H}_{\Lambda_L, J}(\Gamma)$ is the energy of GSP~$\Gamma$ restricted to the volume $\Lambda_L\subset{\mathbb Z}^d$. 
We consider the difference~(\ref{eq:ed}) as a random variable when $\alpha$ and $\beta$ are two GSP's sampled from two restricted metastates.


We may now apply Theorem~5.5 of~\cite{ANSW16}, which in the present context can be expressed as: 

\medskip

\begin{theorem}
\label{thm:flucs}
{\rm (modified from~\cite{ANSW16}):  Consider two infinite-volume GSP's $\alpha$ and $\beta$ chosen from distinct restricted metastates satisfying~(\ref{eq:incong2}), and let~${\cal E}_L(J,\alpha,\beta)$ denote their energy difference as defined in~(\ref{eq:ed}). Then  there is a constant $c>0$ such that the variance of ${\cal E}_L(J,\alpha,\beta)$ under the probability measure $M:=\nu(dJ)\kappa_J(d\omega){\kappa}^{p_1,\delta}_{J,\omega}(d\alpha)\times{\kappa}^{p_2,\delta}_{J,\omega}(d\beta)$ satisfies}
\begin{equation}
\label{eq:flucs}
{\rm Var}_M\Big({\cal E}_L(J,\alpha,\beta)\Big)\ge c\vert\Lambda_L\vert\, .
\end{equation}
\end{theorem}

If $\kappa_J$ is supported on multiple incongruent GSP's, there are two possibilities; the first is that the overlap distribution of the barycenter of $\kappa_J$ is spread over a nonzero interval~\cite{HD07}. Then for any $\alpha$ and $\beta$, there is an $\omega$ for which $q^{(e)}_{\alpha\omega}=p_1$ and $q^{(e)}_{\beta\omega}=p_2$ for some $p_1\ne p_2$. Then $\alpha$ and $\beta$ will belong to different restricted metastates as in~(\ref{eq:incong2}), and the lower bound~(\ref{eq:flucs}) applies.

It could also be the case (though to the best of our knowledge it has not been established) that, as occurs at positive temperature in replica symmetry breaking, there is only a single non-self-overlap value $q_0<1$ at zero temperature.  If so, one uses Construction~1 with $p=q_0$ to build the first metastate~$\kappa^{q_0,\delta}_{J,\omega}$ and Construction~2 with $p=1$ to build the second metastate.  Then any two incongruent GSP's in $\kappa_J$ will belong to incongruent restricted metastates, and one can again apply the lower bound~(\ref{eq:flucs}).
(This procedure can also be used when the overlap distribution of the barycenter of $\kappa_J$ is spread over a nonzero interval.)

There is also an upper bound derived in~\cite{WA90}, which when applied to the situation considered here can be stated as
\begin{equation}
\label{eq:flucsupper}
{\rm Var}_M\Big({\cal E}_L(J,\alpha,\beta)\Big)\le d\vert\Lambda_L\vert\, ,
\end{equation}
where $d>0$ is again positive and independent of the volume. Combining~(\ref{eq:flucs}) and~(\ref{eq:flucsupper}) we therefore have

\begin{theorem}
\label{thm:cd} 
{\rm For any two incongruent GSP's $\alpha$ and $\beta$ in the support of $\kappa_J$, there exist constants $0<c\le d$ such that}
\begin{equation}
\label{eq:bounds}
c\vert\Lambda_L\vert\le{\rm Var}_M\Big({\cal E}_L(J,\alpha,\beta)\Big)\le d\vert\Lambda_L\vert\, .
\end{equation}
\end{theorem}

Theorems \ref{thm:noexist} and \ref{thm:cd} together form the central results of this paper.

\section{Discussion}
\label{sec:discussion}

\subsection{Multiplicity of ground state pairs in two dimensions}
\label{subsec:2D}

These results lead to new insights in two dimensions, where the question of existence of multiple GSP's in the support of~$\kappa_J$ remains open. (A partial result exists for the {\it half\/}-plane, where it was shown for the EA Ising model that there exists only a single GSP~\cite{ADNS10}).

Let $J_L$ denote the set of couplings inside $\Lambda_L$.  It was proved in~\cite{ANSW16} (using a result from~\cite{AW90}) that there exists $c_1>0$ such that the distribution of $M\Bigl[{\cal E}_L(J,\alpha,\beta)\Big\vert J_L\Bigr]/\sqrt{\vert\Lambda_L\vert}$ has the property
\begin{equation}
\label{eq:root}
\liminf_{L\to\infty}\nu\Bigl(\exp t\ \frac{M({\cal E}_L(J,\alpha,\beta)\vert J_L)}{\sqrt{\vert\Lambda_L\vert}}\Bigr)\ge e^{c_1t^2}\, ,
\end{equation}
for all $t$ in any dimension when $\alpha$ and $\beta$ are chosen from incongruent metastates.  In two dimensions $\sqrt{\vert\Lambda_L\vert}$ and $\vert\partial\Lambda_L\vert$ have the same scaling with $L$, so~(\ref{eq:root}) can be replaced by
\begin{equation}
\label{eq:lower}
\liminf_{L\to\infty}\nu\Bigl(\exp t\ \frac{M({\cal E}_L(J,\alpha,\beta)\vert J_L)}{\vert\partial\Lambda_L\vert}\Bigr)\ge e^{c_2t^2}\, .
\end{equation}

In any dimension, an almost sure upper bound on ${\cal E}_L(J,\alpha,\beta)$ can be obtained by decoupling the boundary:
\begin{equation}
\Bigl| {\cal E}_L(J,\alpha,\beta\Bigr| \leq 4 \sum_{e\in \partial \Lambda_L} |J_e|=4 |\partial\Lambda_L|\nu(\vert J_e\vert)\ , \text{ $M$-a.s.}
\end{equation}
which leads to
\begin{equation}
\label{eq:upper}
\limsup_{L\to\infty}\nu\Bigl(\exp t\ \frac{M({\cal E}_L(J,\alpha,\beta)\vert J_L)}{\vert\partial\Lambda_L\vert}\Bigr)\le e^{4t}\, ,
\end{equation}
so the bounds (\ref{eq:lower}) and~(\ref{eq:upper}) are in contradiction for sufficiently large $t$ if incongruent states are present in the support of $\kappa_J$.   This proves

\begin{theorem}
\label{thm:inc}
{\rm For $d=2$, a periodic boundary condition metastate $\kappa_J$ of the Ising EA model at zero temperature cannot be supported on incongruent GSP's.}
\end{theorem}

As mentioned at the beginning of Sect.~\ref{sec:energyflucs}, there is also a result about incongruence in the support of $\kappa_J$:

\medskip

\begin{theorem}
\label{thm:mutual}
{\rm (Newman-Stein~\cite{NS01c} and Theorem~5.2 of~\cite{NS24b}). For any $d\ge 2$, given the Hamiltonian~(\ref{eq:EA}) and a~$\kappa_J$ constructed from it, all non-spin-flip-related pure states in $\kappa_J$ are mutually incongruent.}
\end{theorem}

\medskip

{\bf Remark.} The main result of Theorem~\ref{thm:inc} is the absence of incongruence in $2D$ EA GSP's generated by sequences of volumes with periodic (or antiperiodic~\cite{NS98}) boundary conditions. It leaves open the possibility of the existence of {\it regional congruence\/}~\cite{HF87,FH87}, where distinct GSP's differ by a zero-density interface (as in ferromagnets, for example).  By Theorem~\ref{thm:mutual}, regional congruence cannot appear in ground or pure states generated not only by periodic boundary conditions, but more generally by coupling-independent boundary conditions (including an average over translates if necessary) in any dimension. If regionally congruent ground or pure states do exist, they can be generated only by boundary conditions which are conditioned on the couplings by some as yet unknown procedure. Although of some mathematical interest, they are unlikely to appear in physical systems, and do not correspond to the multiple states predicted by either replica symmetry breaking or chaotic pairs.

\medskip

Combining Theorems \ref{thm:inc} and \ref{thm:mutual} leads to the conclusion:

\medskip

\begin{theorem}
\label{thm:single}
{\rm For $d=2$, a zero-temperature periodic boundary condition metastate $\kappa_J$ for the EA Ising model~(\ref{eq:EA}) is supported on a single ground state pair.}
\end{theorem}

Theorem~\ref{thm:single} asserts that {\it a\/} two-dimensional zero-temperature PBC metastate is supported on a single pair of spin-reversed ground states; this applies as well to an antiperiodic boundary condition metastate constructed along the same (deterministic) subsequence of volumes, which is identical to the corresponding PBC metastate~\cite{NS98}.  

That leaves open the possibility, however, that there may be multiple $2D$ zero-temperature PBC metastates, each supported on a single GSP, but with different metastates supported on different GSP's.   Suppose there exist two PBC metastates $\kappa_J^{(1)}$ and $\kappa_J^{(2)}$, with $\kappa_J^{(1)}$ supported on the single GSP~$\alpha$ and $\kappa_J^{(2)}$ supported on the single GSP~$\beta$ which is distinct from $\alpha$. Then $\alpha$ and $\beta$ are necessarily incongruent, by the same reasoning used in the proof of Theorem~5.2~\cite{NS01c}. Consequently $\kappa_J^{(1)}$ and $\kappa_J^{(2)}$ are incongruent (cf.~Eq.~(\ref{eq:incong2})), and the fluctuations of the energy difference between $\alpha$ and $\beta$ will again be governed by~(\ref{eq:bounds}).  Therefore~(\ref{eq:lower}) and~(\ref{eq:upper}) will again hold, leading to a contradiction.

In~\cite{NS24a,NS24b} it was shown that a {\it positive\/}-temperature PBC metastate in $2D$ also cannot be supported on more than a single pure state pair (numerical evidence~\cite{BY86} strongly suggests that at positive temperature in $2D$ there is in fact only a single pure state). The reasoning above applies to this situation as well, leading to the result:

\medskip

\begin{theorem}
\label{thm:unique}
{\rm The periodic boundary condition metastate $\kappa_J$ for the EA Ising model~(\ref{eq:EA}) in two dimensions is unique (i.e., is the same for all sequences of volumes)  at any temperature, and at zero temperature is supported on a single ground state pair.}
\end{theorem}

\medskip

Theorem~\ref{thm:unique} can be extended to metastates constructed using other coupling-independent boundary conditions, such as all free or all fixed, which can be made translation-covariant by averaging over finite-volume translates~\cite{NS2D01}.  In such cases the extended theorem states that in two dimensions any such metastate is still supported on a single ground state pair, either as an equal mixture in a single Gibbs state or else on the two individual ground states.

This raises the question of whether the single GSP on which (say) the free boundary condition metastate is supported is the same as the single GSP on which the PBC metastate is supported.  Suppose that the PBC metastate $\kappa_J$ is supported on the single GSP $\alpha$ and the free~BC metastate is supported on the single GSP~$\beta$.  If $\alpha$ and $\beta$ are incongruent, then  by~(\ref{eq:incong2}) so are their respective metastates.  By the same line of reasoning that led to Theorem~\ref{thm:flucs}, the energy difference fluctuations ${\cal E}_L(J,\alpha,\beta)$ between $\alpha$ and $\beta$ obey~(\ref{eq:bounds}). Then the argument leading up to Theorem~\ref{thm:inc} shows that $\alpha$ and $\beta$ cannot be incongruent.  On the other hand, the same reasoning that led to Theorem~\ref{thm:mutual} (cf.~\cite{NS01c}) also asserts that $\alpha$ and $\beta$ must be incongruent. This shows that simple coupling-independent boundary condition (periodic, antiperiodic, free, and fixed) EA~metastates in two dimensions are all supported on the same two globally spin-reversed ground states.

\subsection{Excitations with space-filling interfaces}
\label{subsec:RSB}

There are at present four scenarios for the spin glass phase that are consistent both with numerical results and, as far as is currently known, mathematically consistent: replica symmetry breaking (RSB)~\cite{Parisi79,Parisi83,MPSTV84a,MPSTV84b,MPV87,MPRR96,MPRRZ00,NS02,NS03b,Read14,NRS23b}, droplet-scaling~~\cite{Mac84,BM85,BM87,FH86,FH88b}, trivial-nontrivial spin overlap~(TNT)~\cite{KM00,PY00}, and chaotic pairs~\cite{NS96c,NS97,NSBerlin,NS03b}.   A long-standing open question in spin glass theory concerns which (if any) of these pictures is correct, and for which dimensions and temperatures.

The differences among the four pictures at positive temperature are described elsewhere~\cite{NS03a,NS22,NRS23b}, but they also make different predictions at zero temperature. Of the four, RSB and chaotic pairs both predict the existence of many ground states, while scaling-droplet and TNT imply the existence of only a single spin-reversed pair~\cite{FH88b,NS01c,ANS19}.  These differences can all be traced back to different predictions concerning the nature of the {\it interfaces\/} that separate ground states from their lowest-lying long-wavelength excitations. Whether $\kappa_J$ at zero temperature is supported on a single pair of GSP's or multiple incongruent pairs follows from the nature of these interfaces.

An interface between two infinite-volume spin configurations $\tau$ and $\tau'$ is defined to be the set of edges whose associated couplings are satisfied in $\tau$ and unsatisfied in $\tau'$, or vice-versa; they separate regions in which the spins in $\tau$ agree with those in $\tau'$ from regions in which their spins disagree. An interface may consist of a single connected component or multiple disjoint ones, but by the continuity of the coupling distribution, if $\tau$ and $\tau'$ are ground states any such connected component must be infinite in extent.  By definition, incongruent GSP's differ by a space-filling interface.



We now present three methods designed to produce large-lengthscale excitations above the ground state, starting with a method proposed by Palassini and Young~(PY)~\cite{PY00} (see also \cite{PLJY03}), which was one of two papers (the other by Krzakala and Martin~(KM)~\cite{KM00}, which we will return to shortly) which first proposed the TNT picture based on numerical simulations of the EA model in three and four dimensions. The TNT picture proposes that the lowest-energy large-lengthscale excitation above a spin glass ground state (in three and presumably higher dimensions) has $d_s<d$ with energy remaining $O(1)$  over the entire volume in all volumes considered. It was shown in~\cite{NS01c} that if correct the TNT picture predicts a single GSP.

In the PY approach, a perturbation is added to the Hamiltonian~(\ref{eq:EA}) that increases the energy of the ground state so that a different spin configuration could be the new ground state for the perturbed Hamiltonian.  Fixing the coupling configuration~$J$, suppose that in a volume~$\Lambda_L$ with PBC's the GSP is~$\alpha_L$. Then for any spin configuration $\tau_L$ inside $\Lambda_L$, the perturbed energy is given by
\begin{equation}
\label{eq:PY}
{\cal H}^{(PY)}(\tau_L)={\cal H}_L(\tau_L)+ \frac{\epsilon}{\vert \mathbb{E}_L\vert}\sum_{<x,y>\in \mathbb{E}_L}\sigma^{(\alpha_L)}_x\sigma^{(\alpha_L)}_y \sigma^{(\tau_L)}_x\sigma^{(\tau_L)}_y  =-\sum_{<x,y>\in \mathbb{E}_L} J_{xy} \sigma^{(\tau_L)}_x\sigma^{(\tau_L)}_y + \epsilon\ q^{(e)}_{\alpha_L,\tau_L}
\end{equation}
where $\epsilon>0$ is a fixed small parameter.  One then looks for the spin configuration with minimum energy under the perturbed Hamiltonian~(\ref{eq:PY}).  

There are two important things to note about the PY Hamiltonian~(\ref{eq:PY}).  The first is that it raises the energy of the GSP~$\alpha_L$ for the EA Hamiltonian by~$\epsilon$. Because we are looking for the spin configuration~$\alpha^{(PY)}_L$ with minimum energy under~(\ref{eq:PY}), the energy difference $\alpha^{(PY)}_L-\alpha^{(EA)}_L\le\epsilon$ (with equality only if $\alpha^{(PY)}_L$ and $\alpha_L^{(EA)}$ are identical). We are therefore guaranteed that any excited spin configuration uncovered by this method must maintain an energy difference of $O(1)$ with the EA GSP in all volumes.  

The second is that Eq.~(\ref{eq:PY}) is designed to uncover excited spin configurations $\tau_L$ that minimize $q^{(e)}_{\alpha_L,\tau_L}$ and therefore have a maximal interface with the EA GSP.  Simulations and their subsequent analysis led PY to conclude that the lowest-energy excitation in $\Lambda_L$ with $O(1)$ energy above the unperturbed ground state~$\alpha_L$ differs from $\alpha_L$ by an interface of linear size $\ell\sim O(L)$ whose dimension $d_s<d$. 

When comparing these results to predictions from the various proposed scenarios for the spin glass phase, one assumes that this behavior persists for all volumes; this must be the case if PY excitations are to have thermodynamic significance. This extrapolation is done in~\cite{PY00} and similar studies~\cite{MP01} using finite-size scaling arguments.  One can nonetheless arrive at some conclusions about the thermodynamic implications of the PY approach using general arguments, as was done in~\cite{NS01c}.  In particular, we have the following.

\medskip

\begin{theorem}
\label{thm:PY}
{\rm Suppose that the PY procedure is carried out on a sequence of volumes $\Lambda_L$ with $L\to\infty$ (all with PBC's, say).  Then any convergent subsequence of finite-volume spin configurations $\tau_L$ which minimize the energy of ${\cal H}^{(PY)}_L$ is itself an infinite-volume ground state of~(\ref{eq:EA}).}
\end{theorem}

\medskip

{\bf Proof.}  We begin by noting that by standard compactness arguments there must be at least one convergent subsequence of finite-volume spin configurations $\alpha^{(PY)}_L$ which minimize~${\cal H}^{(PY)}_L$; call the resulting infinite-volume spin configuration pair~$\alpha^{(PY)}$. In order to be an infinite-volume GSP, $\alpha^{(PY)}$ must satisfy inequality~(\ref{eq:gs}).  Suppose that in one of the volumes $\Lambda_{L_0}$ along the sequence, the PY GSP $\alpha^{(PY)}_{L_0}$ contains a bounded droplet of spins $D(L_0)$ that violates~(\ref{eq:gs}); i.e., flipping the droplet will lower the energy as computed by~(\ref{eq:gs}) by a fixed amount $e_0>0$ (it is important to note that $\alpha^{(PY)}_{L_0}$ may nonetheless be the PY GSP in $\Lambda_{L_0}$ because it may have minimal edge overlap with the EA GSP $\alpha_L$).  The number of edges $\vert\partial D(L_0)\vert$ in the droplet boundary is bounded from above by $dL_0^d$.

Next consider a second volume $\Lambda_L$ along the sequence with $L\gg L_0$, and ask whether $D(L_0)$ persists.  Assume that the PY GSP $\alpha^{(PY)}_L$ includes the unflipped droplet $D(L_0)$, and let $\tau_L$ denote a spin configuration in $\Lambda_L$ identical to  a second spin configuration $\tau'_L$ but with $D(L_0)$ flipped.  Suppose further that ${\cal H}^{(EA)}(\tau_L)=-\alpha_0$. We then have

\begin{eqnarray}
\label{eq:PYexc}
{\cal H}^{(PY)}(\tau'_L)-{\cal H}^{(PY)}(\tau_L) 
= \Biggl(-\alpha_0+e_0 + \epsilon q^{(e)}_{\alpha^{(PY)}_L\tau'_L}-\Bigl[-\alpha_0+\epsilon\Bigl(q^{(e)}_{\alpha^{(PY)}_L\tau'_L}+\frac{\vert \partial D(L_0)\vert}{\vert \mathbb{E}_L\vert}\Bigr)\Bigr]\Biggr)\nonumber\\
=e_0-\epsilon\frac{\vert \partial D(L_0)\vert}{\vert \mathbb{E}_L\vert}\ge e_0-\epsilon(L_0/L)^d\to e_0\,\,\,\, {\rm as}\,\,\,\, L\to\infty
\end{eqnarray}
so beyond a lengthscale $L\sim L_0(\epsilon/e_0)^{1/d}$ any spin configuration --- including $\alpha^{(PY)}_L$ ---  can lower its PY energy by flipping $D(L_0)$.  Because this is true for any finite droplet, $\alpha^{(PY)}$ is an infinite-volume GSP.~$\diamond$

\medskip

The conclusions of~\cite{PY00} were criticized in~\cite{MP01}, where a similar numerical study was performed. The conclusion of~\cite{MP01} was that in every finite volume, the PY method generates an interface whose boundary scales proportionally to the volume and has energy $O(1)$ above the ground state over the entire volume, and that this behavior persists in the thermodynamic limit.  

What can be said about the properties of the infinite volume ground state and associated excitation generated by this procedure, assuming it persists for all volumes? (The following discussion will be informal, but we note that a more detailed argument would make use of the uniform perturbation metastate discussed in~\cite{NS01c}.)  At this point it is necessary to examine the energy scaling along the interface; i.e., for a large fixed volume, how does the energy of the interface fluctuate as one looks at subvolumes with sides large compared to one but small compared to the entire volume?  The simplest, and (we believe) the most natural possibility is that the energy remains~$O(1)$ in all subvolumes, i.e., does not scale with subvolume size, just as it behaves for the volume as a whole.  

But this cannot be, for the following reason. By Theorem~\ref{thm:PY}, if such excitations above an EA GSP $\alpha$ persist as $L\to\infty$, then a new GSP~$\beta$ is created whose symmetric difference with $\alpha$ is such an interface.  Because the interface is space-filling, $\alpha$ and $\beta$ are incongruent, and therefore~(\ref{eq:bounds}) holds. But in this scenario the energy difference in any restricted volume within the infinite system does not scale with the volume, contradicting Theorem~\ref{thm:cd}. Therefore this possibility cannot be correct.

In principle, it could also be~\cite{Readprivate} that in finite volumes, the energy in its subvolumes {\it does\/} scale in some way with the subvolume, but in such a way that cancellations lead to energy $O(1)$ over the {\it entire\/} volume. Suppose this subvolume scaling is characterized by an exponent $\theta_{\rm sv}$.\footnote{Formally, one can define $\theta_{\rm sv}$ as follows. Take a cube of fixed side $L_0$ centered at the origin inside a volume $\Lambda_L$ with periodic boundary conditions, such that $1\ll L_0\ll L$. Using the PY approach, let $L\to\infty$ and compute the energy difference $\Delta E_{L_0}$ between the excited and ground states inside $\Lambda_{L_0}$. Next take a cube of side $L_1>L_0$ and repeat the process.  Do the same for an infinite sequence of volumes of sides $L_i$, $,i=0,1,2,\ldots$ and define $\theta_{\rm sv}=\lim_{i\to\infty}\ln\Delta E_{L_i}/\ln L_i$. (Note that all of these limits exist along subsequences using standard compactness arguments.)} Once again, for any value of $\theta_{\rm sv}$ smaller than $d/2$, there is a contradiction when one takes the infinite-volume limit. The only remaining possibility that does not contradict Theorem~\ref{thm:cd} is that $\theta_{\rm sv}=d/2$, which to the best of our knowledge does not appear in the literature.

The conclusion is that either an excitation interface whose surface grows proportionally to the volume does not persist for an infinite sequence of finite volumes, or else in the limit the (space-filling) excitation above the ground state has a ``subvolume'' exponent $\theta_{\rm sv}=d/2$.

\bigskip

Two other methods have been proposed to search for large-lengthscale, low-energy (i.e., not diverging as volume increases) excitations.  The first is that of Krzakala and Martin~\cite{KM00}, which appeared simultaneously with the PY paper and came to the same conclusions.  In the KM approach, one considers as before a finite volume $\Lambda_L$ with periodic boundary conditions. Two spins are independently chosen uniformly at random within $\Lambda_L$ and forced to assume a relative orientation opposite to that which they had in the GSP $\sigma_L$.  The resulting excited state, which we again denote by~$\tau_L$, is the lowest energy spin configuration in $\Lambda_L$ in which the chosen pair of spins have the opposite orientation from that in $\sigma_L$.

Once again we consider a sequence of volumes in which a new pair of spins is chosen independently (and uniformly at random) in each separate volume, determining a new $\tau_L$ as before.  As was the case with PY, there will be at least one subsequence in which the $\tau_L$ converge to an infinite-volume spin configuration pair~$\tau$, which itself is a GSP.

To see this, fix a finite volume (or ``window'')~$\Lambda_{L_0}$; as $L\to\infty$ the independently-chosen spins will move outside of $\Lambda_{L_0}$ with probability approaching~one.
Consider a $\Lambda_L$ with $L\gg L_0$, and let $\sigma_1$ and $\sigma_2$ be the two spins chosen independently within $\Lambda_L$, so that  $\tau_L$ is the lowest-energy configuration in $\Lambda_L$ subject to $\sigma_1$ and $\sigma_2$ having the opposite relative orientation to what they had in $\sigma_L$, the EA finite-volume GSP in $\Lambda_L$. In that case~(\ref{eq:gs}) must hold for any contour or surface completely inside $\Lambda_L$ that includes either {\it both\/} or {\it neither\/} of $\sigma_1$ and $\sigma_2$. Because the two chosen spins eventually move outside any finite $L_0$ in the infinite-volume limit, Eq.~(\ref{eq:gs}) becomes satisfied in $\tau_L$ for every closed contour or surface inside {\it any\/} window of fixed size, no matter how large. Therefore any infinite-volume spin configuration $\tau$ which is a convergent subsequence of $\tau_L$'s satisfies the definition of an infinite-volume GSP.

Given that the KM and PY procedures are expected to give similar results, it is natural to ask whether the KM procedure can generate a space-filling interface. Suppose it does, so that the limiting ground states $\sigma$ and $\tau$ are incongruent.  We need to consider how the energy difference fluctuations in ${\cal E}_L(J,\alpha,\beta)$ behave. Unlike the PY case, the energy fluctuations in KM are not necessarily bounded unless the coupling magnitudes are themselves bounded, as would be the case if $\nu(dJ_{xy})$ is, say, a flat distribution in $[-1,1]$. However, we're interested in the case where $\nu(dJ_{xy})$ is Gaussian with mean zero and variance one.

Given a particular $\Lambda_L$ with chosen spins $\sigma_1$ and $\sigma_2$ as before, consider two possible excited spin configurations: $\sigma_L^\prime$ is the configuration identical to $\sigma_L$ except with $\sigma_1$ overturned, and $\sigma_L^{\prime\prime}$ is the configuration identical to $\sigma_L$ except with $\sigma_2$ overturned. Of these, let $\sigma_L^\prime$ have the lower energy. In that case the energy of the KM GSP in $\Lambda_L$ is bounded from above by that of $\sigma_L^\prime$.

In ${\mathbb Z}^d$ each spin has $2d$ neighbors; an upper bound on the energy change caused by flipping a single spin can then be obtained by summing the absolute values of the couplings assigned to the edges attached to that spin, as in~(\ref{eq:ss}). As usual we take the coupling distribution $\nu(dJ_{xy})$ to be Gaussian with mean zero and variance one. Then the distribution of upper bounds for the KM energy difference $\Delta E^{(KM)}_L= E(\tau_L)-E(\sigma_L)$ is the distribution of twice the sum of absolute values of $2d$ random variables $J_i$ chosen from $\nu(dJ_{xy})$. This cannot be written in closed form for finite $d$ but because the $\vert J_i\vert$ are independent tends toward a Gaussian as $d\to\infty$.

For our purposes it is sufficient to find the mean and variance of the random variable $S_{2d}=\sum_{i=1}^{2d}\vert J_i\vert$.  Because means always add, $E[S_{2d}]=2d\sqrt{2/\pi}$, and because the $\vert J_i\vert$ are uncorrelated (in fact independent), the variances also add so that ${\rm Var}[S_{2d}]=2d(1-2/\pi)$, which provides an upper bound in any volume for the variance of $\Delta E^{(KM)}_L= E(\tau_L)-E(\sigma_L)$.  So if the KM method generates a space-filling interface, we arrive at the same conclusions as for the PY method: either the properties found in~\cite{MP01} do not persist for an infinite sequence of volumes, or else the excitation above the ground state has a ``subvolume'' exponent $\theta_{\rm sv}=d/2$.

\bigskip

The third method was proposed in~\cite{NS22}. For each $\Lambda_L$ one independently chooses a bond $b_0$ uniformly at random from the edge set $\mathbb{E}_L$ contained within $\Lambda_L$ and changes the sign of its coupling $J(b_0)$, after which the system is allowed to relax to its lowest-energy spin configuration (which we again denote $\tau_L$).  (Of course, if $J(b_0)$ is unsatisfied in $\sigma_L$, the spin configuration won't change.)  If $J(b_0)=K$ is satisfied in $\sigma_L$, and $J_c^{\sigma_L}(b_0)\in(-K,K)$, then $\tau_L$ is simply $\sigma_L$ after the critical droplet $D(b_0,\sigma_L)$ is flipped.

As in KM, the chosen bond moves outside any fixed window $\Lambda_{L_0}$ as $L\to\infty$, so this process again generates a new GSP $\tau$ along some subsequence of volumes.  If 
the critical droplets generated are space-filling, then a space-filling interface will be generated and will have an energy difference with the GSP of $O(1)$ in any finite volume.  But according to Theorem~\ref{thm:noexist} SFCD's do not exist, so this method will not generate a space-filling interface between $\sigma$ and $\tau$ regardless of subvolume energy scaling.

\bigskip

Before proceeding, we emphasize that these results have no bearing on the accuracy or analysis of any of the numerical simulations in the papers cited above; they apply only to the extrapolation of these results to the thermodynamic limit.  If the results persist for all finite volumes, and the sequence converges to a limiting ground state and excitation, we can ask about the nature of the resulting infinite-volume excitation. We have found that {\it if\/} the resulting excitation has a space-filling interface, then the variance of the fluctuations of the energy difference between the excitation and ground state in a restricted volume $\Lambda_L\subset\mathbb{Z}^d$ diverges as $\vert\Lambda_L\vert$.

In this regard, it is also interesting to note that it has been proposed~\cite{Middleton13,Moore21,VMS24} (see also~\cite{THM11}) that a crossover lengthscale $L^*$ (which is much larger than lengthscales used in current numerical simulations) exists beyond which droplet-scaling theory is the correct description of the zero- or low-temperature phase, and below which RSB-like effects may be dominant.  Verification or refutation of that proposal, however, are beyond the methods used in this paper.

\bigskip

The preceding discussion shows that three different procedures discussed above either do not generate an excitation with a space-filling interface or else, if they do, will generate one whose energy difference fluctuations with the grounds state scale as the square root of the restricted volume.  The question remains whether the same is true for {\it any\/} procedure. We now show that is.  Before doing so, we extend our earlier bound~(\ref{eq:root}), which was sufficient to obtain desired results, to a stronger result due to Aizenman and Wehr~\cite{AW90}:

\medskip

\begin{theorem}
\label{thm:Gaussian}
{\rm (modified from Proposition 6.1 of~\cite{AW90}): Let $E_M(\cdot)$ denote the expectation of a measurable function under $M$, and let ${\cal{\tilde E}}_L(J,\alpha,\beta)=E_M[{\cal E}_L(J,\alpha,\beta)\vert J_L]-E_M[{\cal E}_L(J,\alpha,\beta)]$, where $M$, $\alpha$, and $\beta$ are as in Theorem~\ref{thm:flucs} and $J_L$ denotes the set of couplings inside $\Lambda_L$.  Then the distribution of~${\cal{\tilde E}}_L(J,\alpha,\beta)$ has a Gaussian limit.}
\begin{equation}
\label{eq:AW}
{\cal{\tilde E}}_L(J,\alpha,\beta)/\sqrt{\vert\Lambda_L\vert}\xrightarrow{\text{d} }{\cal N}(0,b)\, ,
\end{equation}
{\rm where ${\cal N}$ is the normal distribution and $b>0$ is a positive finite constant.}
\end{theorem}

\medskip

{\bf Proof.}  It is sufficient to prove Theorem~\ref{thm:Gaussian} by showing that ${\tilde{\cal E}}_L(J,\alpha,\beta)$ satisfies the conditions of Proposition~6.1 in~\cite{AW90}. The quantity  ${\cal{\tilde E}}_L(J,\alpha,\beta)$ itself corresponds to $\Gamma_\Lambda(\eta_\Lambda)$ (with $J_L$ corresponding to $\eta_\Lambda$). In our case the variable $\epsilon$ (or $\epsilon_\alpha$) in~\cite{AW90} equals one and the index $\alpha$ used in~\cite{AW90} is irrelevant here, given that~(\ref{eq:EA}) has only nearest-neighbor pairwise interactions.  The variable $\tau_x$ (or $\tau_{\alpha,x}$) in Proposition~5.2 of~\cite{AW90} corresponds to $(1/2)(\langle\sigma_x\sigma_y\rangle_\alpha -\langle\sigma_x\sigma_y\rangle_\beta)$ and $M_\alpha(T,\{h\},\{\epsilon\})$ of Proposition~6.1 corresponds here to the interface density (if it exists, otherwise the upper density) of the $\alpha-\beta$ space-filling interface. Condition~(iii) of Proposition~5.2 is satisfied because~(\ref{eq:EA}) contains only nearest-neighbor interactions.  Therefore conditions (i), (ii), and (iv) of Proposition~6.1 of~\cite{AW90} are satisfied. The second part of condition~(iii) of Proposition~6.1 applies to positive temperature; however, its purpose is to ensure that a lower bound on the variance of ${\tilde{\cal E}}_L(J,\alpha,\beta)$ is strictly positive, which has already been shown in~Theorem~\ref{thm:flucs}.~$\diamond$

\medskip

Now consider any spin configuration $\beta$ generated by a method such that $\beta$ is the lowest-energy excitation above $\alpha$ in all finite volumes and with a space-filling interface in the limit.  By definition the energy of $\beta$ cannot be lowered by any finite connected droplet of spins, and therefore $\beta$ is itself a ground state, as in the three specific examples above.  The central limit behavior of ${\cal E}(J,\alpha,\beta)/\sqrt{\vert\Lambda_L\vert}$ implies that on large lengthscales fluctuations of ${\tilde{\cal E}}_L(J,\alpha,\beta)$ are again of order~$\sqrt{\vert\Lambda_L\vert}$. 

\bigskip

We conclude with two brief remarks.  The first is that the absence of space-filling critical droplets (cf.~Theorem~\ref{thm:noexist}) may help to simplify other extensions of positive-temperature results to zero temperature (for example, possibly the work on indecomposable metastates in~\cite{Read24}). Of course, critical droplets that flip an infinite number of spins but have zero-density boundaries, which may still create difficulties, have not been ruled out. 

The second relates to a remark made in the final section of~\cite{NS24b} about a potential disconnect between finite-volume and thermodynamical understandings of spin glass stiffness at low temperatures, based on numerical work done in low dimensions~\cite{PY99b,Hartmann99,CBM02,Boettcher04,Boettcher05,WMK14}. With the extension in this paper of the results of~\cite{NS24b} to zero temperature, the discussion in~\cite{NS24b} regarding stiffness applies here as well.






{\bf Acknowledgments.}  We thank A.~van~Enter, J. Machta, and M. Moore for helpful comments on the manuscript, and N.~Read for useful discussions that improved and clarified the discussion in Section~6.  We also thank two anonymous referees for their useful comments and questions.

\bigskip

{\bf Data Availability Statement.} Data sharing not applicable --- no new data generated.

\bigskip

{\bf Conflict of Interest Statement.} The authors have no relevant financial or non-financial interests to disclose or competing interests to declare that are relevant to the content of this article.

\bigskip

\appendix
\section{Continuity of flexibility: An example}
\label{app:flex}

Consider the space-filling critical droplet of the bond $b_0$ in a specific GSP $\sigma$ with variable associated coupling $J(b_0)$.   All other
couplings are held fixed throughout. Because $\sigma$ is fixed, the critical value of $J(b_0)$ will simply be denoted $J_c$. For $J(b_0)>J_c$ the ground state is $\sigma^>$; when $J(b_0)<J_c$ the ground state
is $\sigma^<$; they are related by a flip of the SFCD of $b_0$.

We will begin with $J(b_0)>J_{\rm upper}$ and then lower $J(b_0)$ to $J_c$.  When $J(b_0)>J_{\rm upper}$ or else is sufficiently above $J_c$, suppose that a bond $b_1\in\partial D(b_0,\sigma)$ and
$b_2\notin\partial D(b_0,\sigma)$ but $b_1\in\partial D(b_2)$.  See Fig.~\ref{fig:1}.

\begin{figure}[h]
\centering
\includegraphics[scale=0.4]{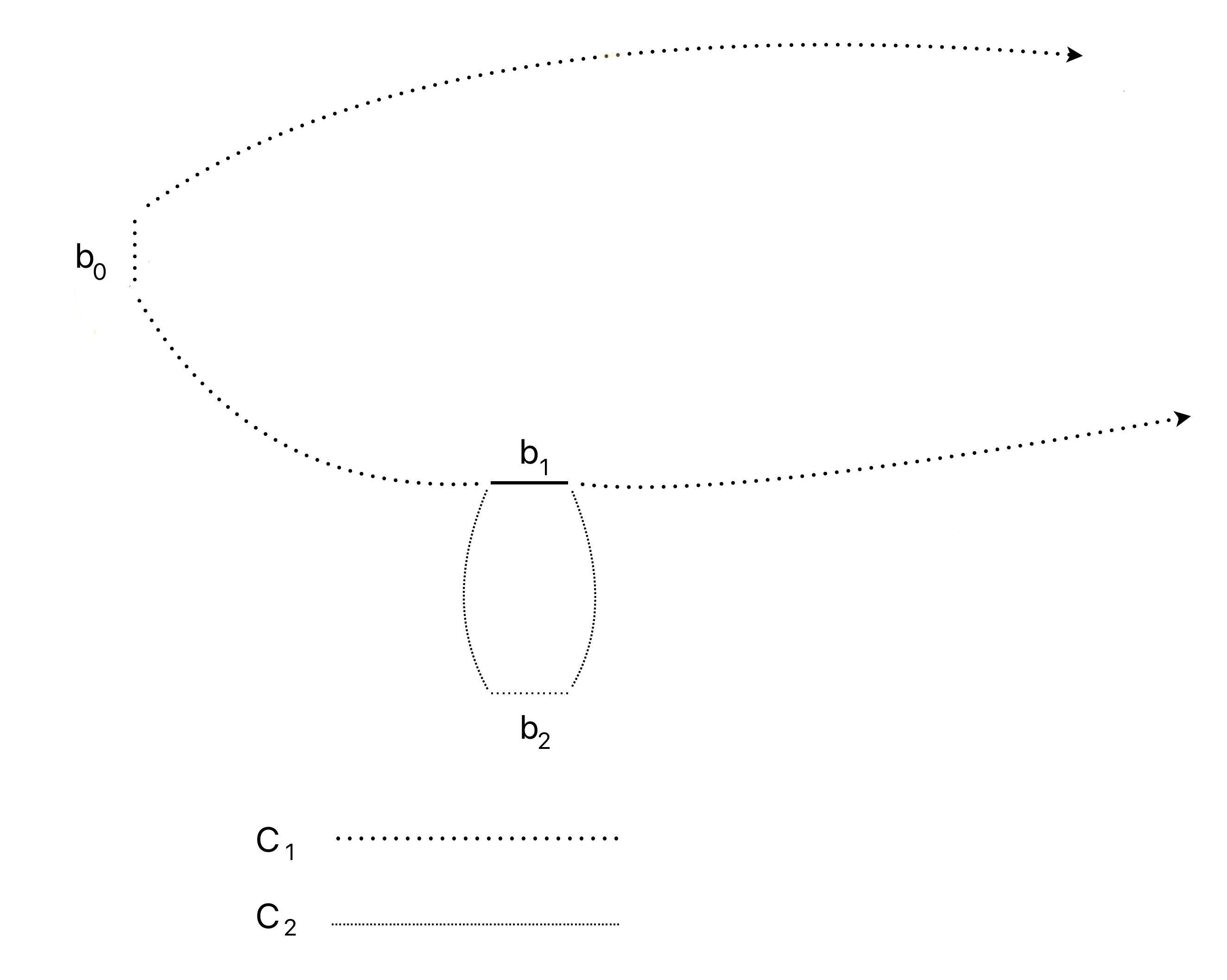}
\caption{Sketch of SFCD of $b_0$ discussed in text. Here $C_1$ refers to the critical droplet boundary of $b_0$ with the single bond $b_1$ removed and $C_2$ refers to the critical droplet boundary of
$b_2$ with the single bond $b_1$ removed.}
\label{fig:1}
\end{figure}

Referring to Fig.~\ref{fig:1}, $C_1$ and $C_2$ are defined so that  $b_1\notin C_1$ and $b_1\notin C_2$ and when $J(b_0)$ is sufficiently above $J_c$, ${\partial D}(b_0)=C_1\cup b_1$ and ${\partial D}(b_2)=C_2\cup b_1$.   We will require that for {\it any\/} value of $J(b_0)$, $b_2$ does not lie in $\partial D(b_0,\sigma)$. We begin by considering the case where $\partial D(b_2,\sigma)$
contains a single bond in $\partial D(b_0,\sigma)$, and will then generalize to multiple bonds.

Let $J_1$ denote the (fixed) coupling value associated with $b_1$. We first consider the case where $J_1$ is unsatisfied in $\sigma^>$.

\medskip

Case 1 ($J_1$ is unsatisfied in $\sigma^>$):  We start with $J(b_0)$ above $J_{\rm upper}$ so that $\partial D(b_2,\sigma)$, the critical droplet boundary of $b_2$, is the union of $b_1$ with $C_2$ (see diagram).
Let the energy of the critical droplet of $b_2$ equal $c$, which by definition must be positive; write this as $E(D(b_2))=c>0$. Because the energy contribution of $b_1$ in $\sigma$
is $-\vert J_1\vert$, we have $E(C_2)=c+\vert J_1\vert$. 

Now lower $J(b_0)$ just past its critical value to $J_c^-$, so $E(D(b_0,\sigma))=0^-$. We are now in the ground state $\sigma^<$ in which the SFCD has flipped and $J_1$ is now satisfied, so given that $E(D(b_0,\sigma))=0^-$, we now have $E(C_1)=-\vert J_1\vert$. If the critical droplet of $b_2$ still included $b_1$, its energy would be $c+2\vert J_1\vert$. But $C_2\cup b_1$ is no longer the critical droplet of $b_2$: the lowest energy droplet whose boundary includes $b_2$ is now the droplet $C_1\cup C_2$, excluding $b_1$. Its energy is $E(D(b_2))=E(C_1)+E(C_2)= -\vert J_1\vert +(c+\vert J_1\vert)=c$, so the flexibility of $b_2$ varies smoothly when $J(b_0)$ passes through $J_c$.

\medskip

Case 2 ($J_1$ is satisfied in $\sigma$): This case is slightly more involved. Here the energy contribution of $b_1$ in $\sigma$ is $\vert J_1\vert$, so we have $E(C_2)=c-\vert J_1\vert$.
But we also require that $b_2$ never become a bond in $\partial D(b_0,\sigma)$, so we must have $c>2\vert J_1\vert$; otherwise, the critical droplet of $b_0$ will deform to include $C_2$ and $b_2$ and
exclude $b_1$.

Now begin lowering $J(b_0$) from $J_{\rm upper}$. At some point still well above $J_c$, $E(C_1)$ will become less than $\vert J_1\vert$, so the critical droplet of $b_2$ will deform to exclude $b_1$ and include $C_1$: i.e., $\partial D(b_2) =  C_2\cup C_1$.    (The critical droplet of $b_0$ is unchanged and still includes $b_1$, as long as $J(b_0)>J_c$.)

When $J(b_0)=J_c^+$, the critical droplet energy of $b_2$ is $E(D(b_2)) = E(C_2) +E(C_1) = (c-\vert J_1\vert)-\vert J_1\vert=c-2\vert J_1\vert$. When $J(b_0)$ passes through $J_c$, i.e., $J(b_0)=J_c^-$, the critical droplet of $b_2$ will again change to include $b_1$, which is now unsatisfied and whose energy contribution to $\partial D(b_2)$ is now $-\vert J_1\vert$ (while $E(C_1)=\vert J_1\vert$). So the critical droplet energy of $b_2$ in $\sigma^<$ with $J(b_0)$ just below $J_c$ is $E(\partial D(b_2)) = E(C_2) +E(b_1)= (c-\vert J_1\vert) -\vert J_1\vert = c-2J_1$, and again there is no flexibility jump at $J_c$.

\bigskip

This argument can be extended to the case where a bond not in the critical droplet of $b_0$ has more than one edge in $\partial D(b_0,\sigma)$. To simplify notation, let $b_N$ denote a bond not in $\partial D(b_0,\sigma)$ but whose critical droplet contains bonds $b_1,b_2,\ldots b_n$ (with corresponding coupling values $J_1,J_2\ldots J_n$) all of which are in $\partial D(b_0,\sigma)$.  Formally, $\partial D(b_0,\sigma)\cap\partial D(b_N,\sigma)=\{b_1,b_2,\ldots b_n\}$. 

Next let $C_1$ denote the surface of the critical droplet of $b_0$ minus $\{b_1,b_2,\ldots b_n\}$ and let $C_2$ denote the surface of the critical droplet of $b_N$ minus $\{b_1,b_2,\ldots b_n\}$. We note that $C_1\cup C_2$ also represents a closed surface in the dual lattice. When $J(b_0)=J_c^+$, there are two cases to consider: $E(J_1)+E(J_2)+\ldots E(J_n) = E_T > 0$ and $E_T < 0$. Now the same arguments go through as for the single-bond case.

\bibliography{refs.bib}

\end{document}